\documentclass[12pt]{article}
\usepackage[dvips]{graphicx}

\newcommand{\GeV}{\ensuremath{\mathrm{Ge\kern -0.12em V}}}
\newcommand{\TeV}{\ensuremath{\mathrm{Te\kern -0.12em V}}}

\begin{document}
\begin{center}

{\Large\bf Exploring the LHC Landscape with Dileptons}
\vspace{1cm}

{\sc 
Dimitri Bourilkov
}

{\small bourilkov@mailaps.org
}

{\sl Physics Department, University of Florida, P.O. Box 118440\\
Gainesville, FL 32611, USA}

\end{center}

\begin{abstract}
The dilepton decay channels provide clean signatures and are an ideal
hunting ground for high mass resonant, like Z', or non-resonant, like
contact interactions or extra dimensions, searches at the LHC. The
production of high invariant mass opposite sign lepton pairs in
proton-proton collisions in the Standard Model is dominated by the
Drell-Yan process. In addition to this photon or Z exchange mediated
mechanism, photons radiated by the incoming protons can collide and
produce lepton pairs. In this paper detailed calculations of the
Drell-Yan process at next-to-next-to-leading order in QCD and
next-to-leading order in the electroweak corrections, augmented with
the photon-induced effects, are presented in the typical acceptance of
a multi-purpose LHC detector at center of mass energy 13~\TeV.
Estimates of the expected backgrounds for new physics searches are
provided for dilepton invariant masses up to the LHC kinematic limit.
\end{abstract}

\section{Introduction}

The Standard Model (SM) of particle physics has ruled the collider
scene for decades. Varied and very detailed tests of the SM have
confirmed its validity at the available energies. Searches for new
physics phenomena beyond the SM in dilepton (electron or muon) final
states provide clean signatures and have been a mainstay of the quest
strategy. Resonant or non-resonant effects have been searched for
extensively at hadron colliders like the LHC, see
e.g~\cite{Chatrchyan:2011wq,Khachatryan:2014fba,Khachatryan:2016zqb,CMS:2016abv,Aad:2014cka,Aad:2014wca,Aaboud:2016cth,ATLAS:2016cyf},
and in the cleaner environment of lepton colliders at lower energies,
see e.g.~\cite{Schael:2013ita,Bourilkov:2001tp,Bourilkov:1998sr}.

The backgrounds for new high mass resonant (like Z') or non-resonant
(like contact interactions or extra dimensions) effects are dominated
by the Drell-Yan (DY) process of opposite sign lepton pair production,
mediated through photon or Z exchange from the initial partons in the
incoming protons. With the rapidly increasing accumulated luminosity
at the LHC precise estimations of the background are a key ingredient
of the dilepton searches, especially of the non-resonant variety. To
reach the required precision, calculations at next-to-leading order
(NLO) and next-to-next-to-leading order (NNLO) in Quantum
Chromodynamics (QCD) are needed. NNLO cross section calculations
reduce the dependence of the results on the renormalization and
factorization scale choices to the couple of percent level, as
expected when enough orders are included in the calculation.

The QCD calculations depend on the parton density functions (PDF) of
the protons. The following modern
PDFs~\cite{Dulat:2015mca,Harland-Lang:2014zoa,Ball:2014uwa,Manohar:2016nzj}
are used in this study. The PDF uncertainties are estimated using the
latest PDF4LHC~\cite{Butterworth:2015oua} prescriptions. In most cases
the CT14 PDF set, the PDF4LHC15 set (which is an average of the CT14,
MMHT14 and NNPDF3.0 PDF sets) or the LUXqed\_plus\_PDF4LHC15 set are
used. Additional PDF sets are utilized for special purposes. The
reweighting technique~\cite{Bourilkov:2006cj} for PDF uncertainties is
used.

Beside the QCD effects, electroweak (EWK) effects become very
important at LHC energies. Technically (and as implemented in various
calculation or simulation tools), they fall in two classes:
\begin{enumerate}
 \item Quantum Electrodynamics (QED) only effects: Final State
       Radiation (FSR), Initial State Radiation (ISR) and their
       interference
 \item Pure Weak corrections: vertex, WW and ZZ box, and self-energy
       contributions.
\end{enumerate}

In addition to the DY process, lepton pairs can be produced in
gamma-gamma collisions, where photons radiated by the incoming protons
collide. To calculate this process, usually labeled photon-initiated
(PI) background in various searches, we need parton density functions
including the photon component. Quantum Electrodynamics
introduces corrections to the parton evolution: photon parton
distributions $\gamma(x,Q^2)$ are present for the proton (neutron),
and part of the proton (or neutron) momentum is carried by the
photons. The PDF depends on the parton momentum fraction - Bjorken
$x$, and the momentum transfer $Q^2$. In this study the modern photon
PDFs~\cite{Manohar:2016nzj,Schmidt:2015zda} are used.
In Drell-Yan, W and Z production at the LHC the photon contribution
is suppressed by a factor $\mathcal{O}(\alpha / \alpha_s)$ compared to
the canonical quark-antiquark contribution.

For an up-to-date paper on the issues involved in precision studies of
the DY process, see e.g.~\cite{Alioli:2016fum}, which concentrates on
W and Z boson production. In contrast, the focus of this paper is on
the high mass search region up to the LHC kinematic limit - a not so
well explored area of phase space.

\section{Setup}

The calculations for the Drell-Yan process and the photon-induced
background are carried out with the program
\mbox{{\tt FEWZ}~\cite{Li:2012wna}}.
The $G_{\mu}$ scheme with the W mass, the Z mass and the Fermi
constant $G_{\mu}$ (measured in muon decay) as input parameters
besides the fermion masses is used. 
The strong coupling is set to $\alpha_{s}(M_Z)\ =\ 0.118$.
The PDFs considered in this study are CT14, NNPDF30, PDF4LHC15 and
LUXqed\_plus\_PDF4LHC15, as provided by the LHAPDF libraries version 5
or 6~\cite{Whalley:2005nh,Bourilkov:2003kk,Buckley:2014ana}.

Full electroweak corrections at next-to-leading order (NLO) are
computed (the flag {\tt EW control = 0} is used). QCD effects are
computed at next-to-leading order (NLO) and next-to-next-to-leading
order (NNLO). When the PI background is added to the Drell-Yan cross
section we label the results \mbox{DY+PI}.

Calculations for dielectrons or dimuons in the acceptance of a generic
general purpose LHC experiment are presented: both outgoing leptons
are required to have pseudorapidity $|\eta| < 2.4$. Relatively hard
cuts suitable for searches at high invariant masses extending to the
multi-$\TeV$ region are used - the transverse momenta for both leptons
have to satisfy $p_T > 50\ \GeV$.

An important difference between the two channels is the treatment of
photons located close in space to the leptons. They can originate from
Final State Radiation, or from unrelated sources, e.g. the
copious decays of $\pi^0$ in a hadron collider environment. 
The variable $\Delta R$ is used to measure ``closeness'':
\begin{equation}
\Delta R = \sqrt{(\Delta \eta)^2 + (\Delta \varphi)^2},
\end{equation}
where $\eta$ and $\varphi$ are the pseudorapidities and azimuthal
angles of the lepton and the photon.

In the case of electrons the electromagnetic calorimeters of the
experiments provide a natural ``integration'' of the energies of
close-by photons with the electron energies. As a result the invariant
mass reconstructed from the electron energies is closer to the
electron pair mass before FSR. In the case of muons where the
transverse momenta can be measured both in the central trackers, and
in the outer muon detectors after the photons are absorbed, there is
no such effect. Two lepton definitions as available in {\tt FEWZ} are
used in the calculations:
\begin{enumerate}
 \item ``Dressed'' electrons: photons within $\Delta R < 0.1$, and 
       with $|\eta| < 2.5$ and transverse momentum $p_T > 0.5\ \GeV$
       are included in the electron energies
 \item ``Bare'' muons: photon energy is not included.
\end{enumerate}

The CT14qed set is introduced in~\cite{Schmidt:2015zda} . The initial
photon distribution is defined by the initial photon momentum fraction
\begin{equation}
p_0^{\gamma} = \int_{0}^{1} xf_{\gamma/p}(x,Q_0) dx
\end{equation}
at scale $Q_0\ =\ $1.295~\GeV.

From the ZEUS data on deep inelastic scattering with isolated
photons~\cite{Chekanov:2009dq} the photon PDF is constrained to
$p_0^{\gamma} \le 0.14\%$.  The constraint can be improved in
measurements where the photon contribution is enhanced by selecting
exclusive dimuon pair production in elastic, single dissociative and
double dissociative pp
collisions~\cite{Chatrchyan:2011ci,Khachatryan:2016mud}. These CMS
measurements are used in~\cite{Ababekri:2016kkj} to update the CT14qed
analysis:
\begin{eqnarray}
p_0^{\gamma} \le 0.09\%\ at\ 68\%\ CL \\
p_0^{\gamma} \le 0.13\%\ at\ 90\%\ CL
\end{eqnarray}
consistent with the ZEUS data analysis.

Recent developments determine the photon PDFs to even higher
precision. The importance of taking into account both the coherent
(photon emission from the proton as a whole) and non-coherent
components in the initial photon distribution is highlighted
in~\cite{Harland-Lang:2016kog}, and a well determined photon PDF is
obtained for high mass lepton pair production. A model-independent
approach~\cite{Manohar:2016nzj} looks in great detail at the photon
distribution, and constrains it quite well with electron-proton
scattering data. The result is the LUXqed\_plus\_PDF4LHC15 photon PDF
set.

\newpage
\section{Cross Section Calculations}

\vspace{0.34cm}
{\bf EWK corrections}

\begin{figure}[ht]
\centerline{\resizebox{0.88\textwidth}{10.5cm}{\includegraphics{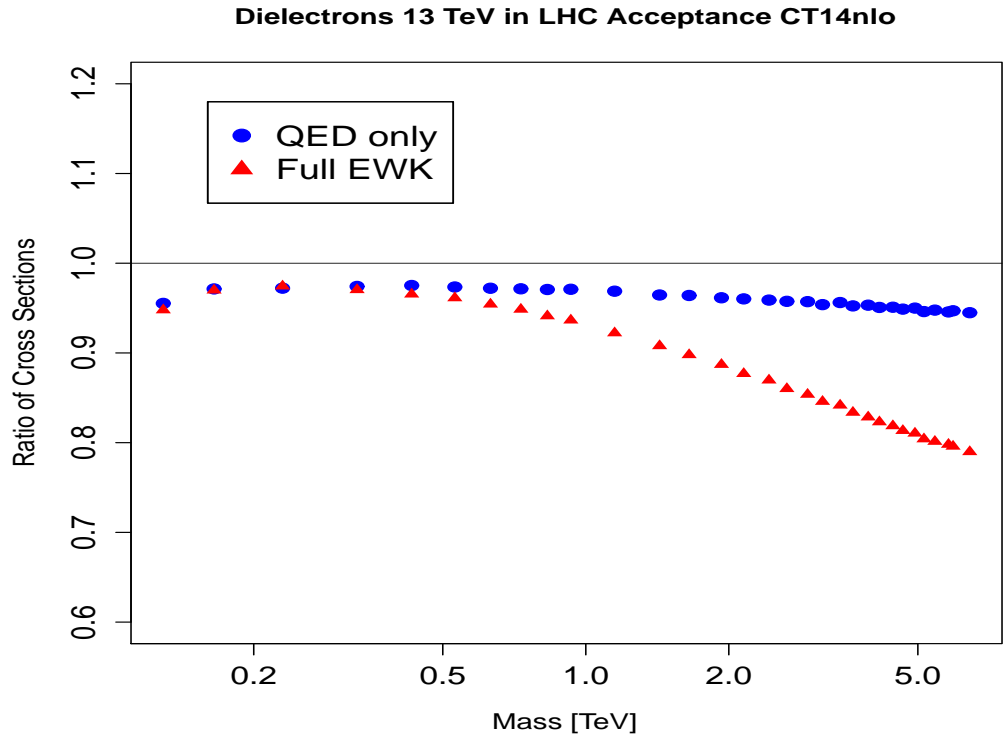}}}
\centerline{\resizebox{0.88\textwidth}{10.5cm}{\includegraphics{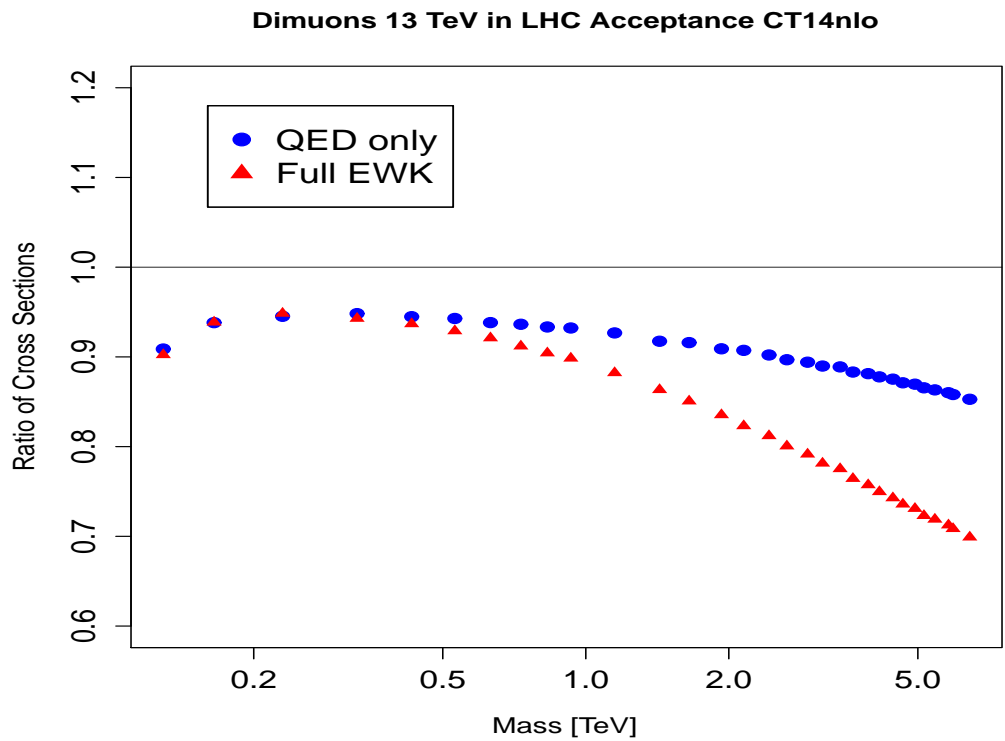}}}
\vspace*{-3pt}
\caption{Top: Electroweak corrections for the dielectron channel:
         QED only and full EWK corrections.
         Bottom: Electroweak corrections for the dimuon channel:
         QED only and full EWK corrections.}
\label{fig:fig1}
\end{figure}

In Figure~\ref{fig:fig1} the cross section ratios at NLO in QCD and
electroweak corrections are shown. We compare three sets of
calculations:
\begin{enumerate}
 \item QCD at NLO, no EWK corrections
 \item QCD at NLO and QED corrections: FSR, ISR and their
       interference; Weak corrections off
 \item QCD at NLO and full EWK corrections.
\end{enumerate}

As can be seen the pure QED corrections are bigger in the dimuon
channel due to the use of ``bare'' muons. The full EWK corrections are
very important at high invariant masses: they reduce the cross section
by more than 20\% in the dielectron channel and by up to 30 \% in the
dimuon channel. In some Monte-Carlo simulations only QED effects are
taken into account. Clearly the inclusion of full EWK effects is key
for successful comparisons to data. The importance of complete EWK
corrections for LHC at 14 $\TeV$ was recognized
early~\cite{Haywood:1999qg}. The results presented here agree well
with calculations in this exploratory study. Higher order corrections
of mixed QCD-EWK type $\mathcal{O}(\alpha_s\alpha)$ are ignored in {\tt
FEWZ}. They are generally less important~\cite{Dittmaier:2015rxo}
except for high precision studies.

\vspace{0.34cm}
{\bf PDF uncertainties}

\begin{figure}[ht]
\centerline{\resizebox{0.88\textwidth}{10.5cm}{\includegraphics{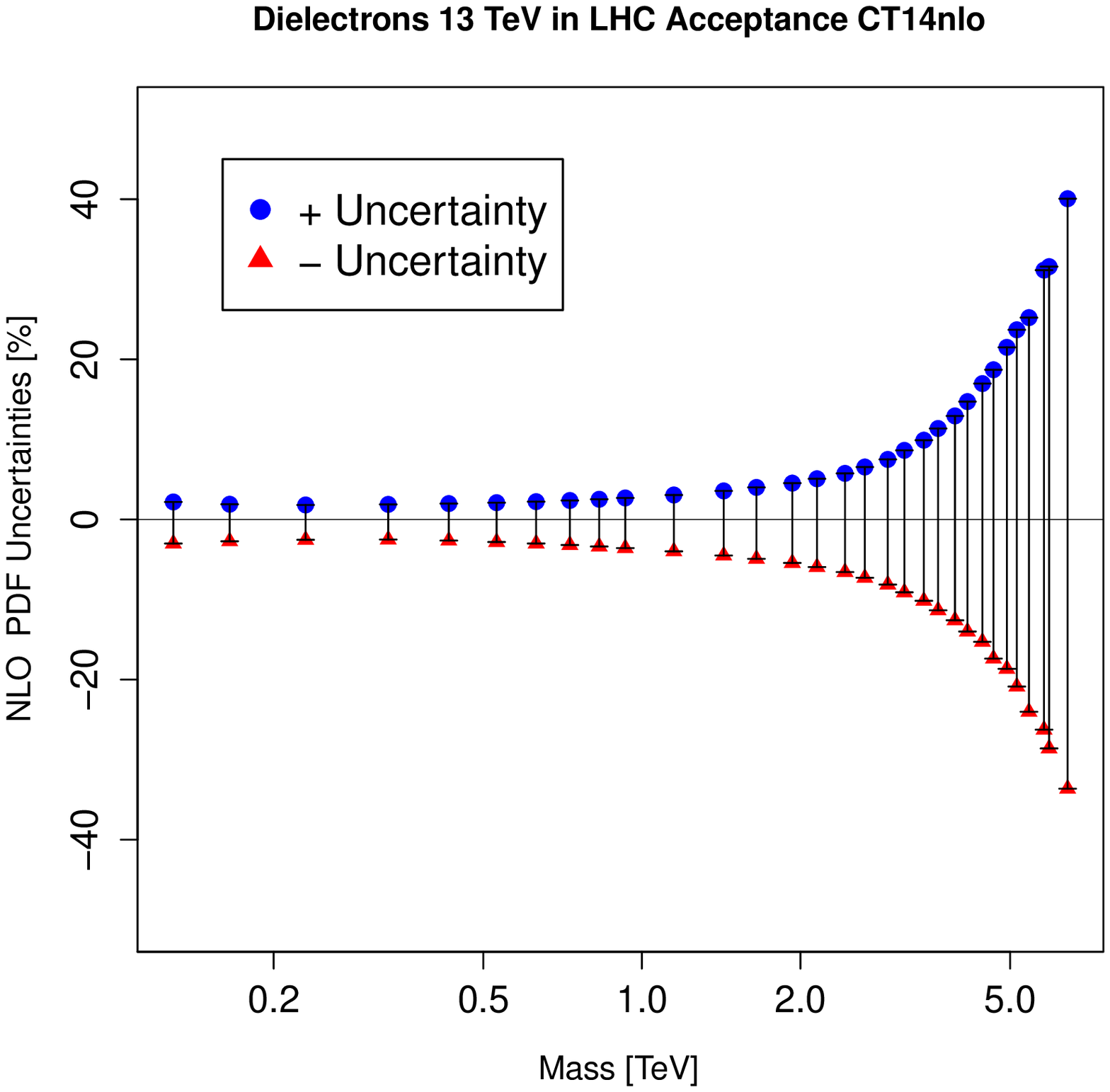}}}
\centerline{\resizebox{0.88\textwidth}{10.5cm}{\includegraphics{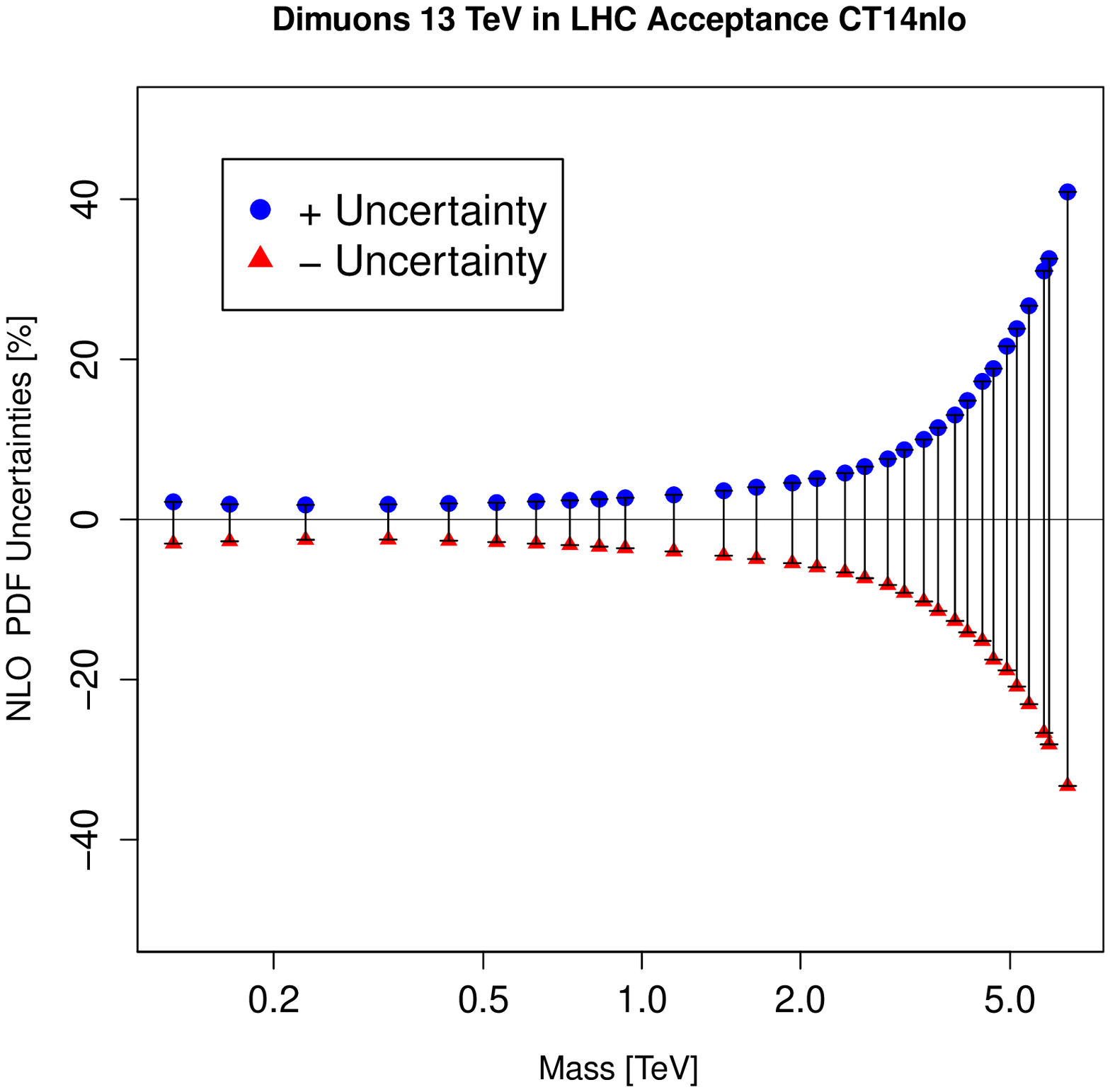}}}
\vspace*{-3pt}
\caption{Top: NLO PDF uncertainties for the dielectron channel.
         Bottom: NLO PDF uncertainties for the dimuon channel.}
\label{fig:fig2}
\end{figure}

In Figure~\ref{fig:fig2} the PDF uncertainties for the NLO cross
sections (QCD and full EWK), using the CT14nlo PDF, are shown as a
function of mass. They become sizable above 2~$\TeV$ and approach
$\sim\pm$~40\% for masses above 5~\TeV. As discussed later, the
PDF uncertainties are dominant at high mass.

\vspace{0.34cm}
{\bf PDF choice}

\begin{figure}[ht]
\centerline{\resizebox{0.88\textwidth}{10.5cm}{\includegraphics{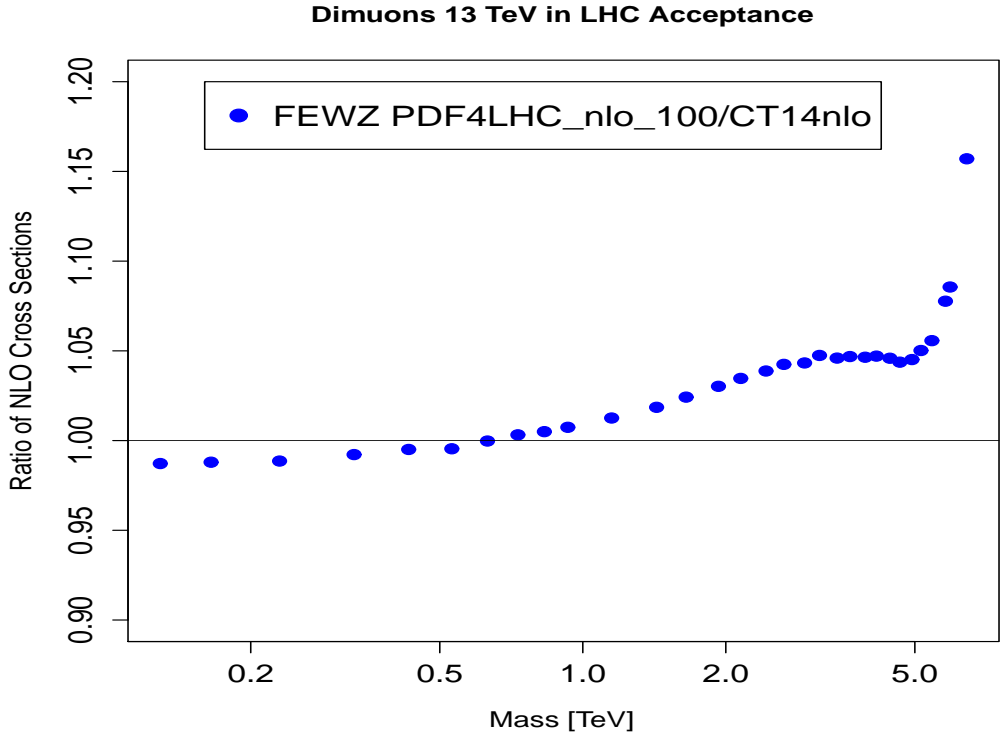}}}
\centerline{\resizebox{0.88\textwidth}{10.5cm}{\includegraphics{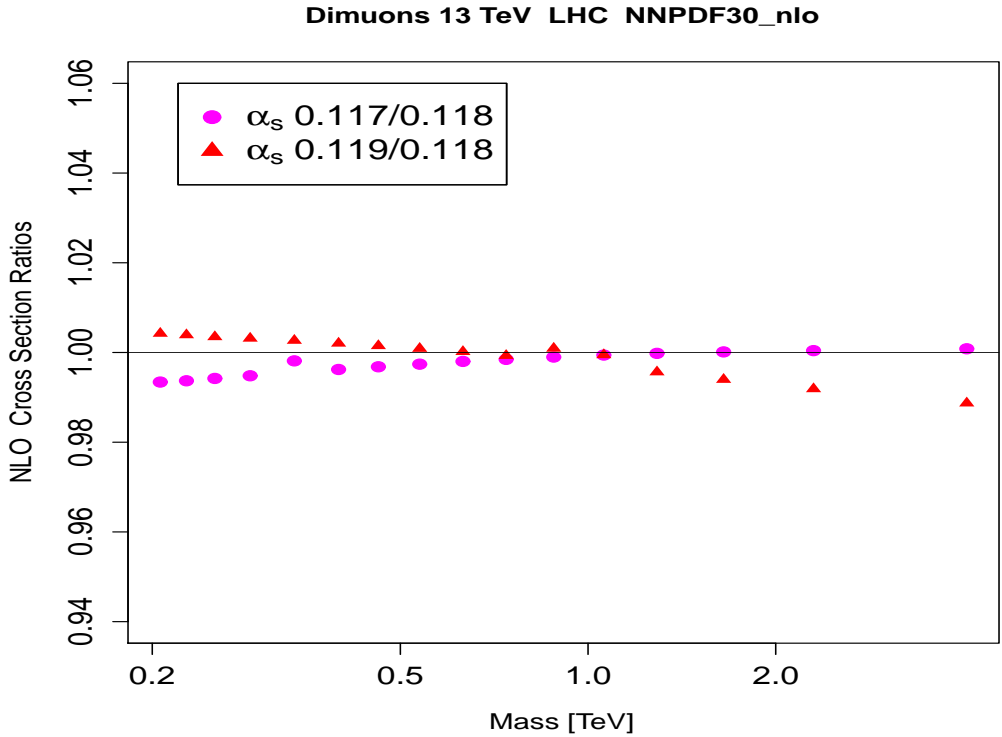}}}
\vspace*{-3pt}
\caption{A: Dependence of the cross sections on the choice of
         PDFs for the dimuon channel. The CT14nlo and
         PDF4LHC\_nlo\_100 sets are compared.
         B: Dependence of the cross sections on variations in the
         value of the strong coupling constant $\alpha_s$
         for the dimuon channel.}
\label{fig:fig2a}
\end{figure}

The dependence of the cross sections on the choice of PDF is displayed
in Figure~\ref{fig:fig2a}A for the CT14nlo and PDF4LHC\_nlo\_100
sets. For masses below 5~$\TeV$ the variation is below 5\%, reaching
$\sim$~15\% for the highest masses. The variation from one is covered
by the PDF uncertainties, and could become important only for very
high integrated luminosities - as shown in the next section even for
100 fb$^{-1}$ no events from SM sources are expected above 5~$\TeV$.

\vspace{0.34cm}
{\bf $\alpha_s$ dependence}

In Figure~\ref{fig:fig2a}B the dependence of the cross sections on
variations in the value of the strong coupling constant $\alpha_s$ is
examined. For variations larger then the current error from the world
average~\cite{Agashe:2014kda}:
\begin{equation}
\alpha_{s}(M_Z)\ =\ 0.1185 \pm 0.0006,
\end{equation}
the effects are below 1\% for the whole mass range under study.

\vspace{0.34cm}
{\bf PI background}

\begin{figure}[ht]
\centerline{\resizebox{0.88\textwidth}{10.5cm}{\includegraphics{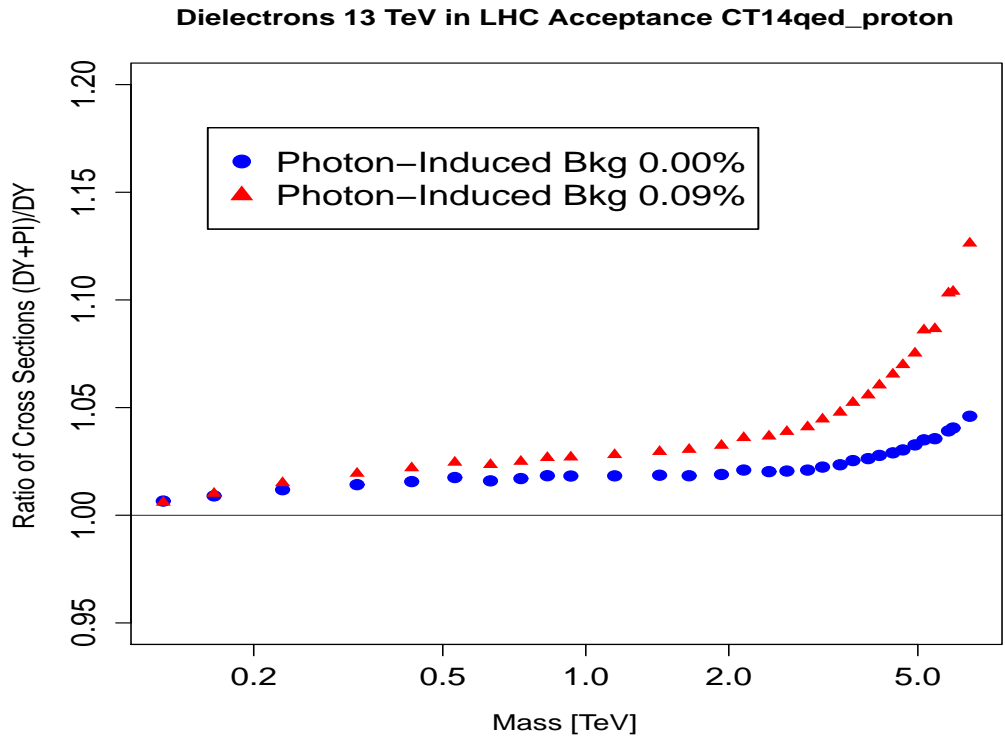}}}
\centerline{\resizebox{0.88\textwidth}{10.5cm}{\includegraphics{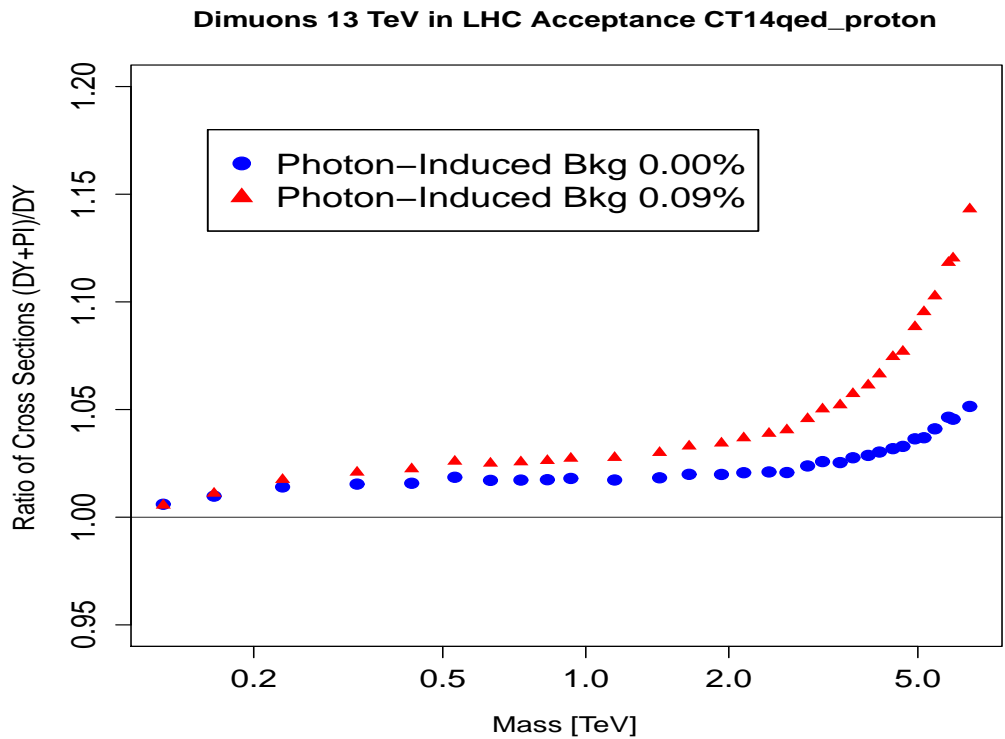}}}
\vspace*{-3pt}
\caption{Top: Photon-induced background for the dielectron channel.
         Bottom: Photon-induced background for the dimuon channel.
         The cross section ratios (DY+PI)/DY are displayed.}
\label{fig:fig3}
\end{figure}

The effect of photon-initiated lepton pair production on the cross
sections is shown in Figure~\ref{fig:fig3}, where the ratios
(DY+PI)/DY are displayed. The CT14qed\_proton PDF is
used. It includes the photon contribution. The initial photon momentum
fraction $p_0^{\gamma}$, which is a free parameter, is varied between
0.00\% and 0.09\% as discussed earlier, and the two limiting cases are
displayed. Their difference is used as one standard deviation in the
first event rates estimate.

\begin{figure}[ht]
\centerline{\resizebox{0.88\textwidth}{10.5cm}{\includegraphics{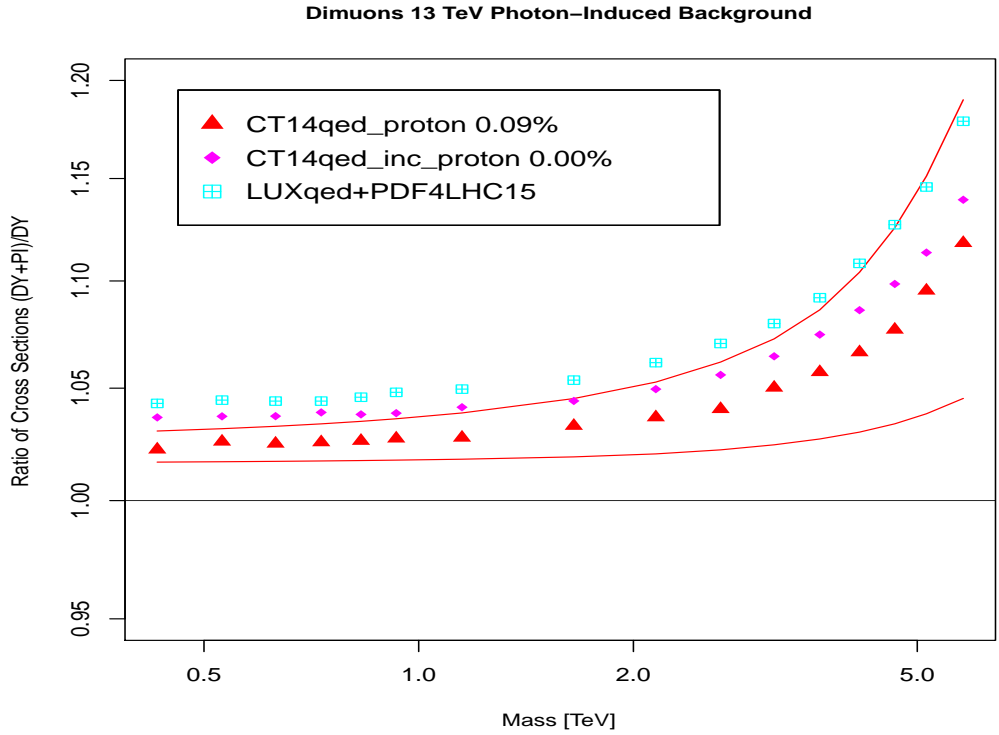}}}
\centerline{\resizebox{0.88\textwidth}{10.5cm}{\includegraphics{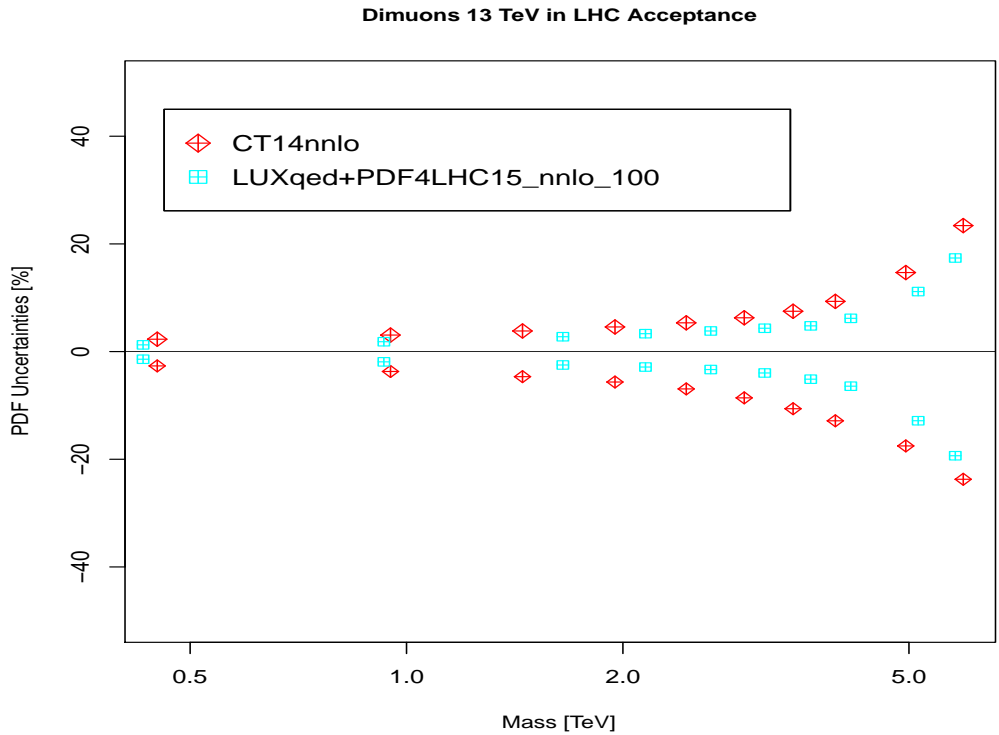}}}
\vspace*{-3pt}
\caption{Top: Photon-induced background for the dimuon channel.
         The cross section ratios (DY+PI)/DY are displayed. The lines
         represent the one $\sigma$ uncertainty band for CT14qed\_proton,
         as described in the text.
         Bottom: NNLO PDF uncertainties for the dimuon channel with
         CT14nnlo and LUXqed\_plus\_PDF4LHC15\_nnlo\_100.}
\label{fig:fig3a}
\end{figure}

The recent photon PDFs, publicly available as of this writing, are
compared in Figure~\ref{fig:fig3a} for the dimuon channel. The
CT14qed\_inc\_proton PDF is based on CT14nlo with the initial photon
PDF defined by the sum of the inelastic photon PDF and the elastic
photon PDF, obtained from the equivalent photon approximation. The
inclusion of the elastic component enhances the PI contribution. The
results from the LUXqed\_plus\_PDF4LHC15 set show the highest PI
effects~\footnote {The results for lepton pair production with
LUXqed\_plus\_PDF4LHC15 are close to the results
from~\cite{Harland-Lang:2016kog}.}. This set is used for the second
estimate of event rates. It benefits also from smaller PDF
uncertainties at NNLO, see Figure~\ref{fig:fig3a}.

The photon-initiated effects are generally small, not above the 5\%
level for masses of up to $\sim$~2~$\TeV$, and can reach
$\sim$~15--20\% above 5~$\TeV$ if the photon momentum fraction is
taken to be 0.09\% or if the LUXqed\_plus\_PDF4LHC15 set is used. The
one $\sigma$ band defined above for CT14qed\_proton PDF is also shown
in Figure~\ref{fig:fig3a}. The predictions from the
CT14qed\_inc\_proton PDF are inside this band for masses above
1.5~$\TeV$, while for the LUXqed\_plus\_PDF4LHC15 set they fall inside
the band only for the highest masses. The newest photon PDFs predict
somewhat stronger PI effects in the lower half of the mass range under
study.

The results presented here agree well with the results
from~\cite{Bourilkov:2016qum} based on the pioneering
MRST2004qed~\cite{Martin:2004dh} PDF set. The predictions from the
freely parametrized NNPDF photon PDF~\cite{Ball:2013hta}, also
analyzed in~\cite{Bourilkov:2016qum}, and the newer
version~\cite{Bertone:2016ume}, have large PDF uncertainties in the
most interesting search region\footnote{Subsequent studies are
confirming these conclusions, see e.g.~\cite{Accomando:2016tah}.}.

\vspace{0.34cm}
{\bf NNLO/NLO K functions and NNLO PDF uncertainties}

\begin{figure}[ht]
\centerline{\resizebox{0.88\textwidth}{10.5cm}{\includegraphics{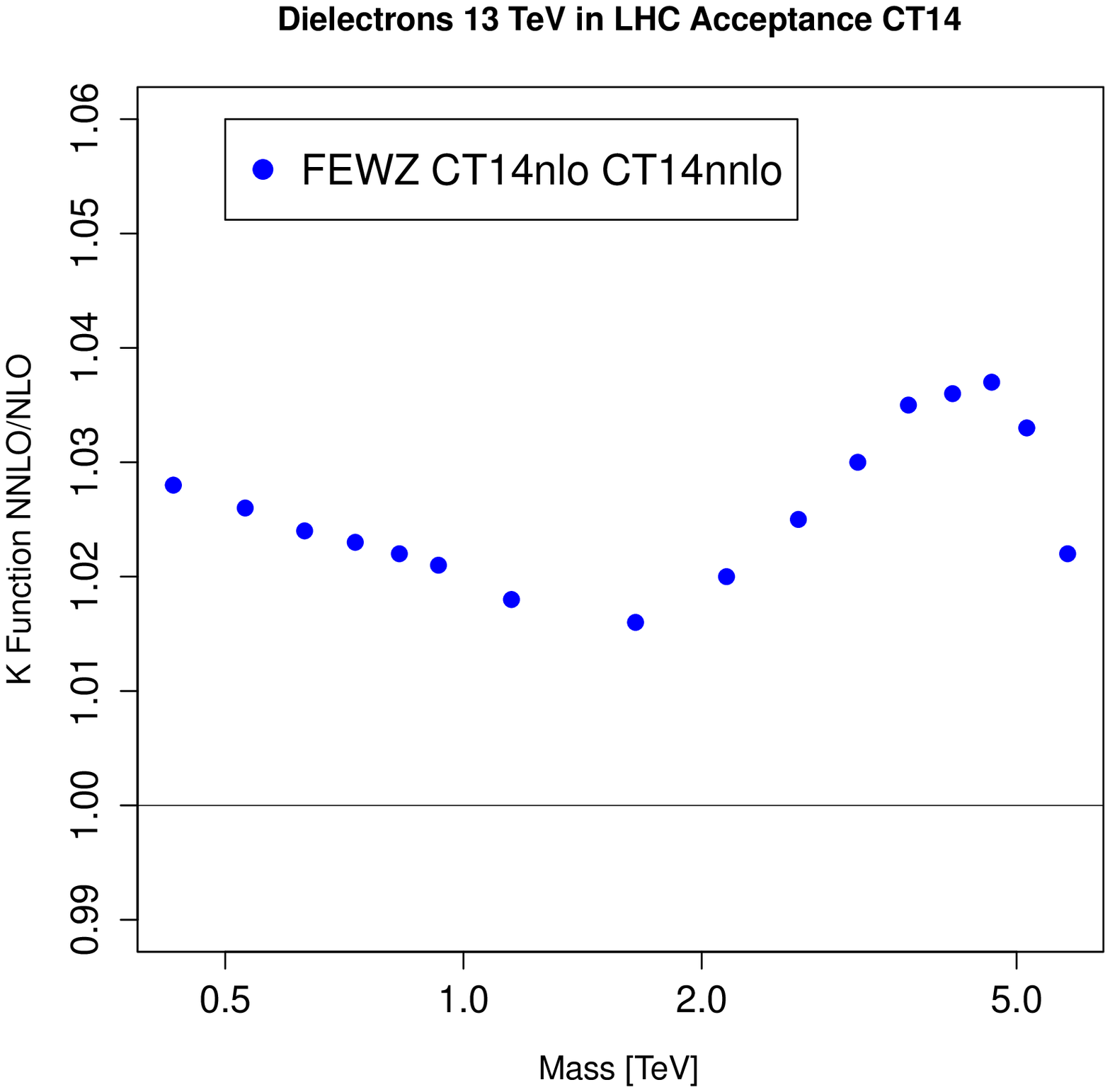}}}
\centerline{\resizebox{0.88\textwidth}{10.5cm}{\includegraphics{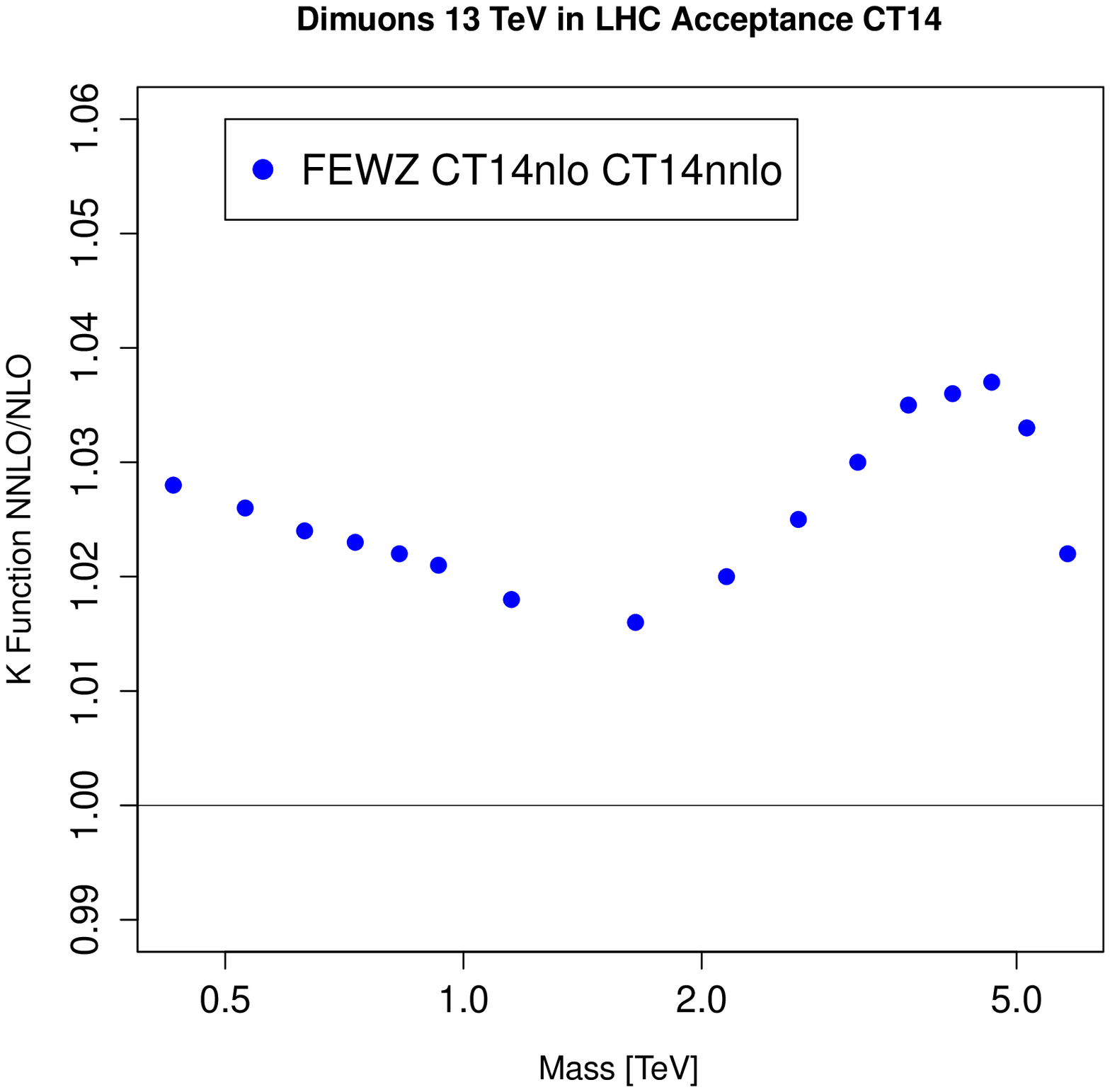}}}
\vspace*{-3pt}
\caption{Top: K function NNLO/NLO for the dielectron channel.
         Bottom: K function NNLO/NLO for the dimuon channel.}
\label{fig:fig4}
\end{figure}

In Figure~\ref{fig:fig4} the cross section ratios for calculations at
NNLO or NLO in QCD, the so called K functions, are shown for
dielectrons and dimuons. In all cases full EWK corrections are
included. The K functions are below 1.04 for the whole mass range
under study, so the NNLO QCD effects are quite small. The K functions
exhibit a complex behavior as a function of mass.

\begin{figure}[ht]
\centerline{\resizebox{0.88\textwidth}{10.5cm}{\includegraphics{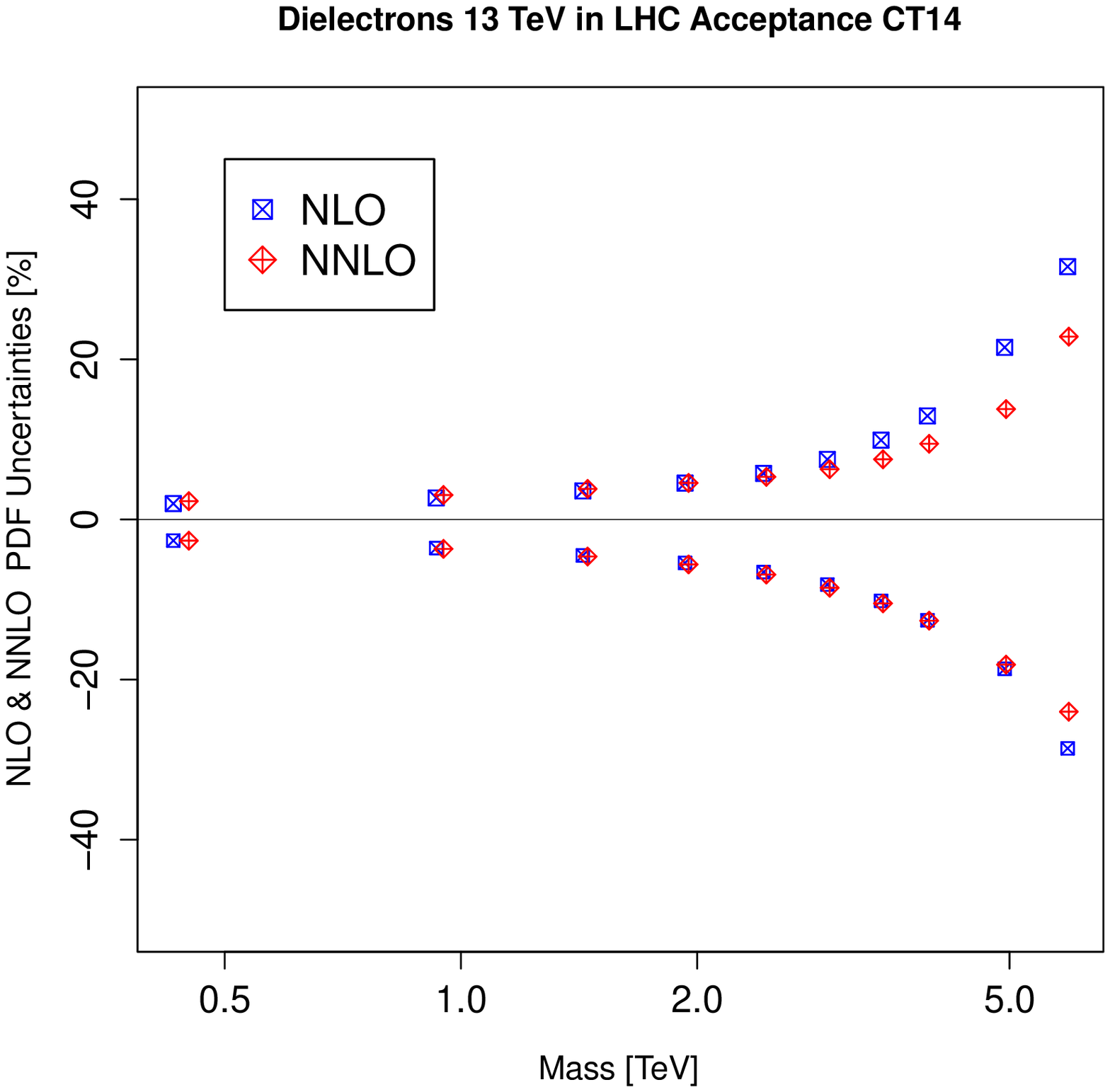}}}
\centerline{\resizebox{0.88\textwidth}{10.5cm}{\includegraphics{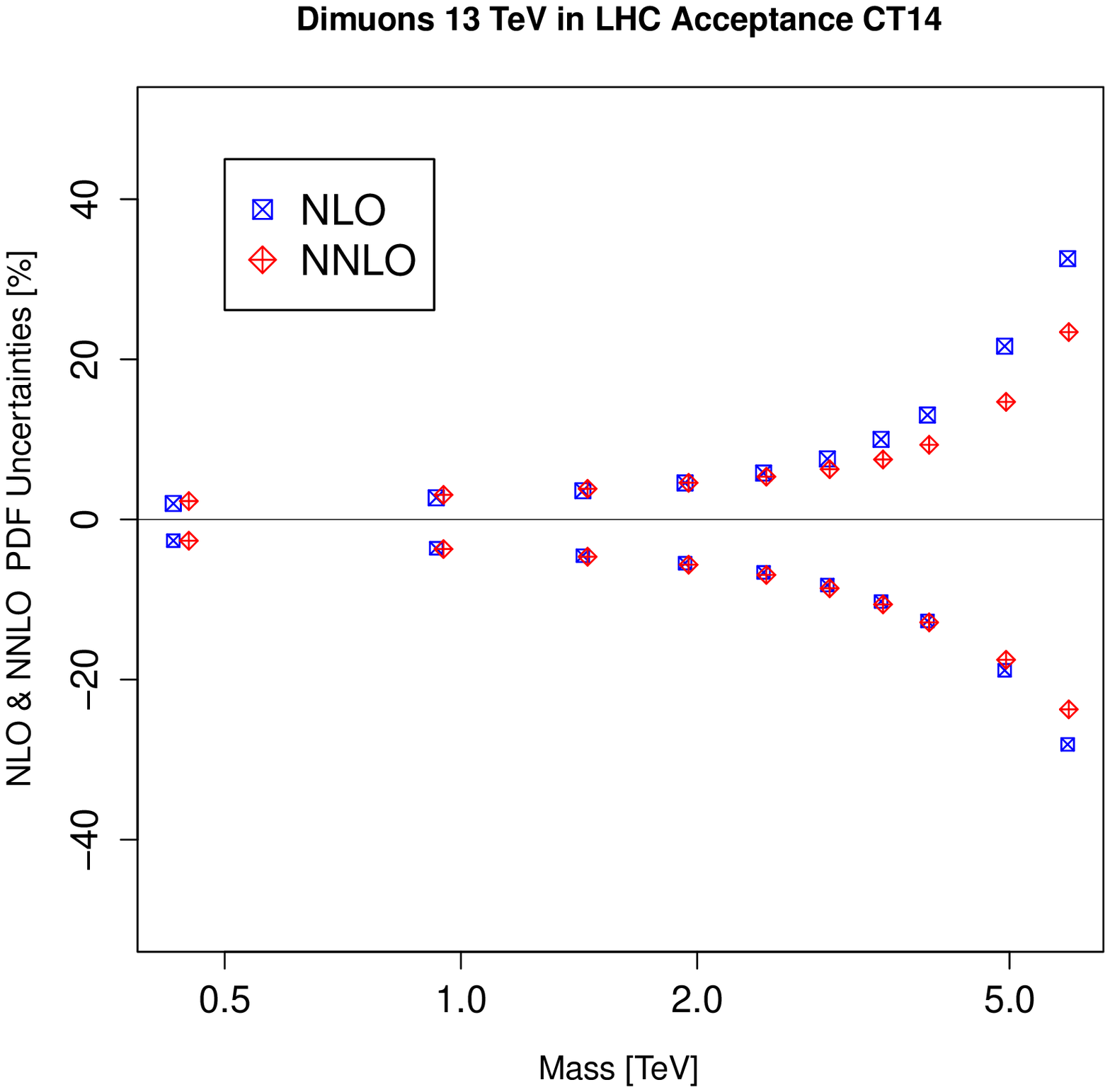}}}
\vspace*{-3pt}
\caption{Top: NLO and NNLO PDF uncertainties for the dielectron channel.
         Bottom: NLO and NNLO PDF uncertainties for the dimuon channel.}
\label{fig:fig5}
\end{figure}

Figure~\ref{fig:fig5} compares the PDF uncertainties for NLO and NNLO
cross sections (QCD and full EWK), using the order-matched CT14nlo and
CT14nnlo PDFs. The good news is that the uncertainties are reduced at
NNLO.

\vspace{0.34cm}
{\bf Scale dependence}

\begin{figure}[ht]
\centerline{\resizebox{0.88\textwidth}{10.5cm}{\includegraphics{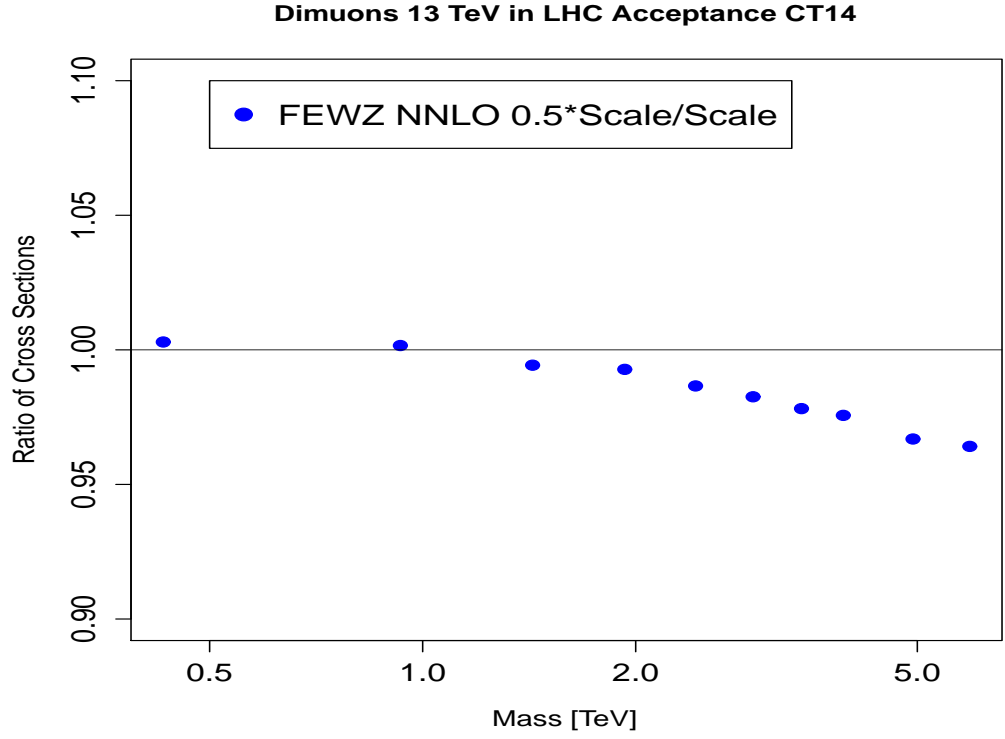}}}
\vspace*{-3pt}
\caption{Dependence of the cross sections on the choice of
         renormalization and factorization scales
         for the dimuon channel.}
\label{fig:fig5a}
\end{figure}

If enough orders are included in the perturbative expansion, the
results should not depend on the choice of renormalization and
factorization scales. Traditionally this effect is estimated by
varying the scales by a factor of two around the nominal scale, which
for Drell-Yan is taken to be the mass of the outgoing dilepton
system. An example of such variation for NNLO calculations is shown in
Figure~\ref{fig:fig5a}. The cross sections change by less than 3\%
below masses of 5~$\TeV$. For the highest masses the variation reaches
3.6\%. At NNLO in QCD the calculations are precise enough to serve the
needs of searches for new phenomena.

\newpage
\vspace{0.34cm}
{\bf All combined}

\begin{figure}[ht]
\centerline{\resizebox{0.88\textwidth}{10.5cm}{\includegraphics{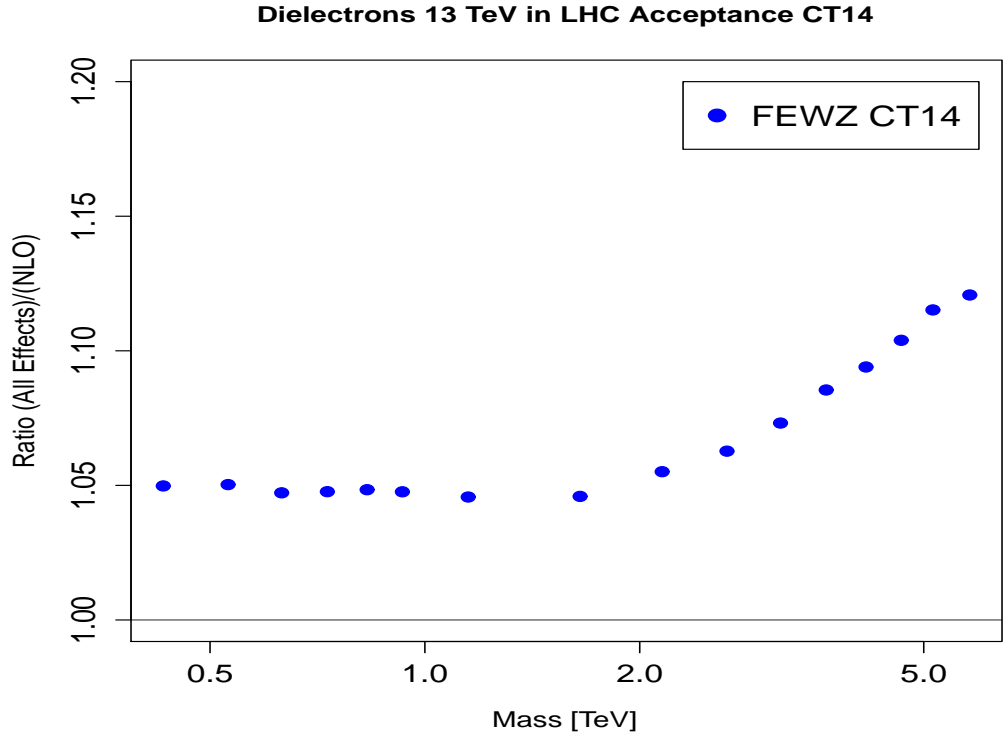}}}
\centerline{\resizebox{0.88\textwidth}{10.5cm}{\includegraphics{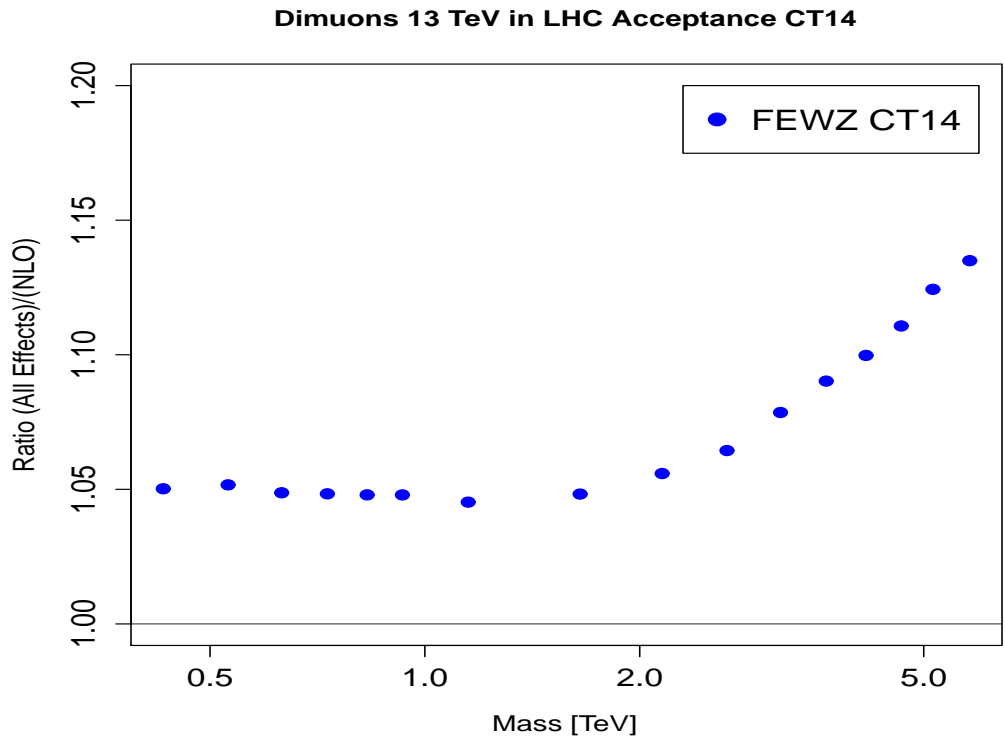}}}
\vspace*{-3pt}
\caption{Top: Ratio of cross sections including all effects (NNLO and PI)
         to NLO cross sections for the dielectron channel.
         Bottom: Ratio of cross sections including all effects (NNLO and PI)
         to NLO cross sections for the dimuon channel.}
\label{fig:fig6}
\end{figure}

In Figure~\ref{fig:fig6} the K functions and photon-induced effects
are combined. The ratios of NNLO cross sections including PI
contributions to NLO cross sections (QCD and full EWK) are shown,
using the CT14 PDF sets. Results with all effects taken into account
will be discussed in the rest of this paper.

\section{Differential Cross Sections and Event Yields}

\begin{figure}[ht]
\centerline{\resizebox{0.88\textwidth}{10.5cm}{\includegraphics{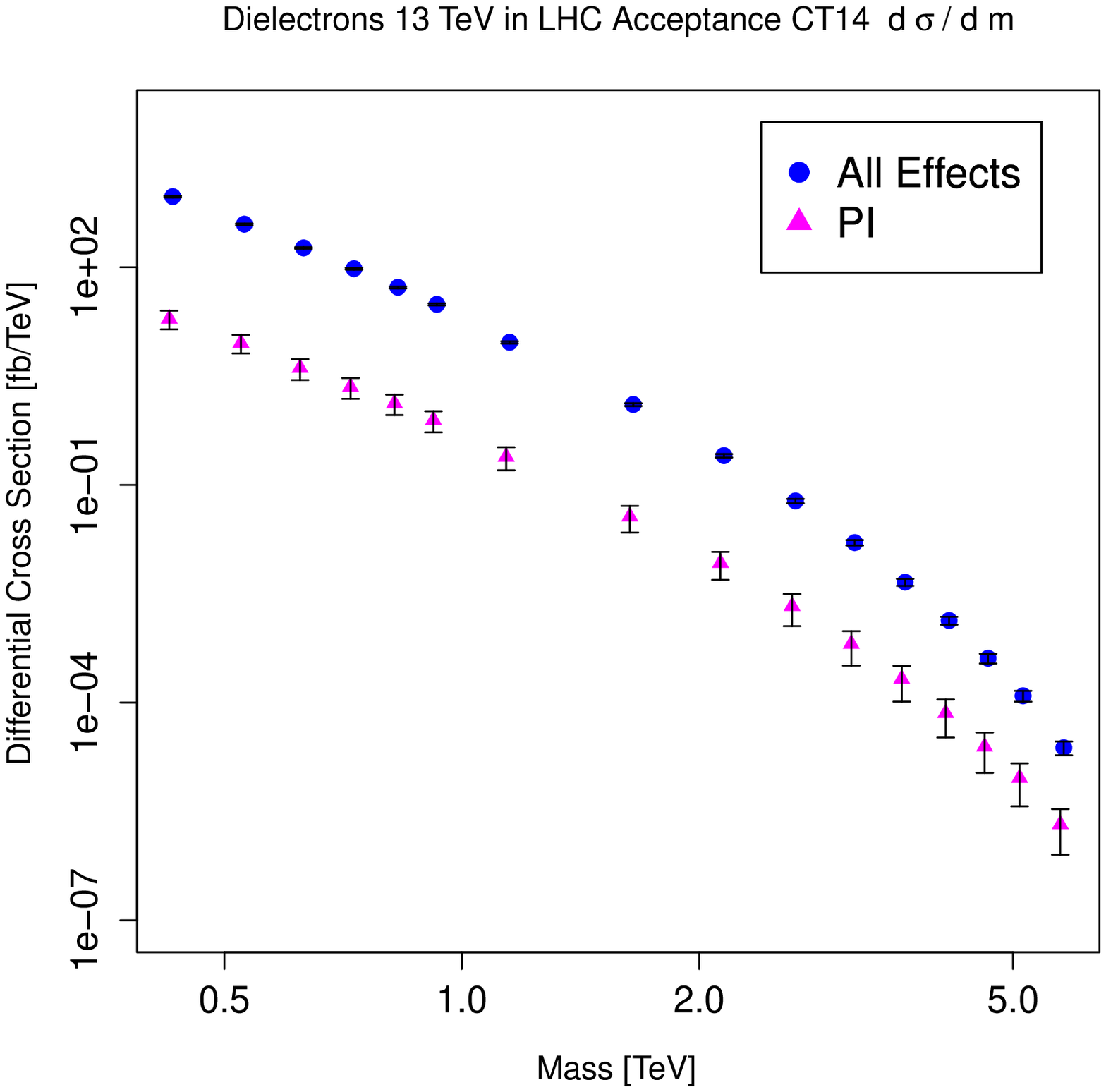}}}
\centerline{\resizebox{0.88\textwidth}{10.5cm}{\includegraphics{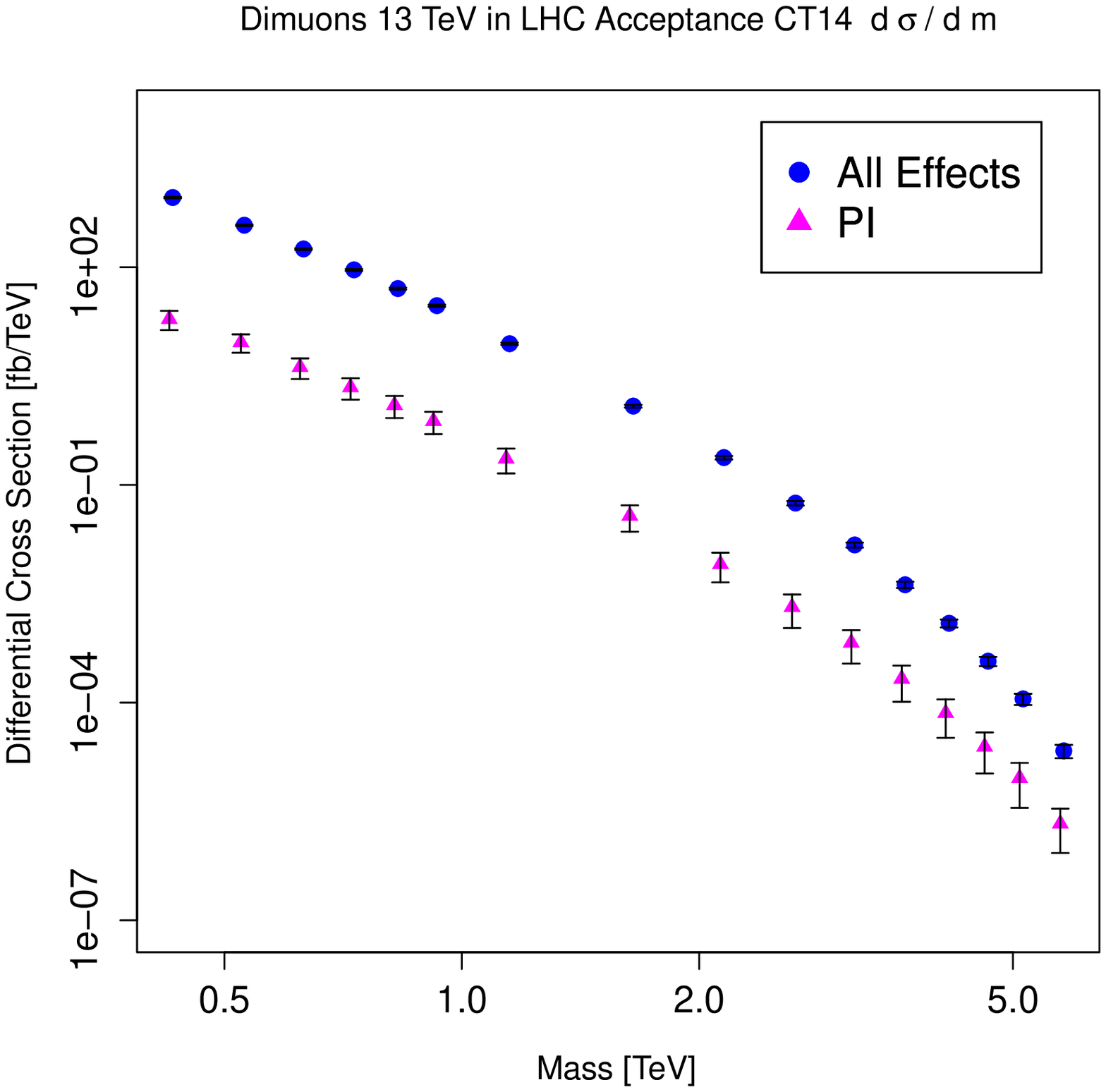}}}
\vspace*{-3pt}
\caption{Top: Differential cross section including all effects (NNLO and PI)
         for the dielectron channel.
         Bottom: Differential cross section including all effects (NNLO and PI)
         for the dimuon channel.}
\label{fig:fig7}
\end{figure}

The differential cross sections as function of mass are shown in
Figure~\ref{fig:fig7} for the dielectron and dimuon channels. The CT14
PDFs are used for the first set of results. The uncertainty of the
photon-induced background, as defined in the previous section, can be
seen on the plots. As this background is much smaller than the
Drell-Yan contribution, the impact on the combined cross section is
minor.

\begin{figure}[ht]
\centerline{\resizebox{0.88\textwidth}{10.5cm}{\includegraphics{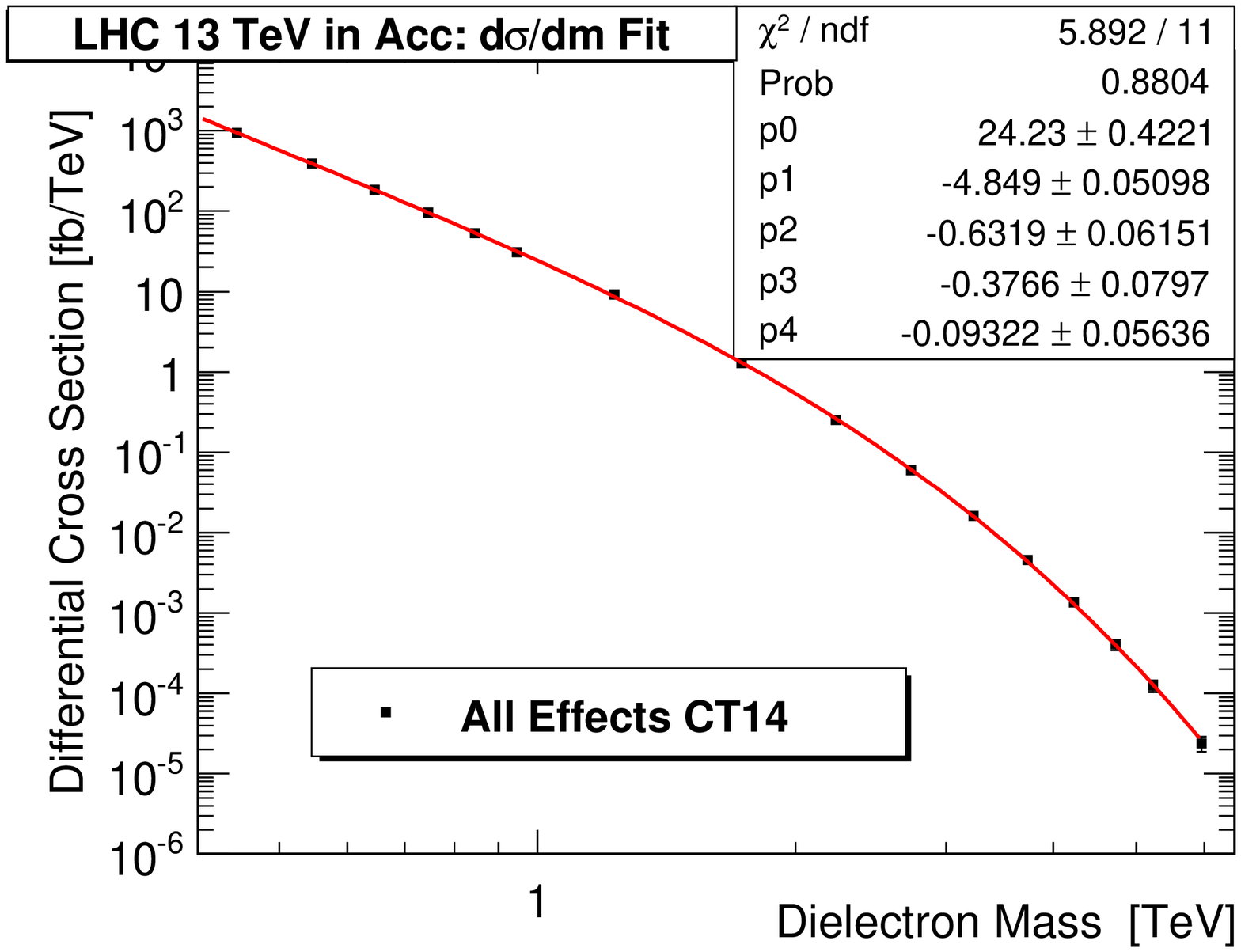}}}
\centerline{\resizebox{0.88\textwidth}{10.5cm}{\includegraphics{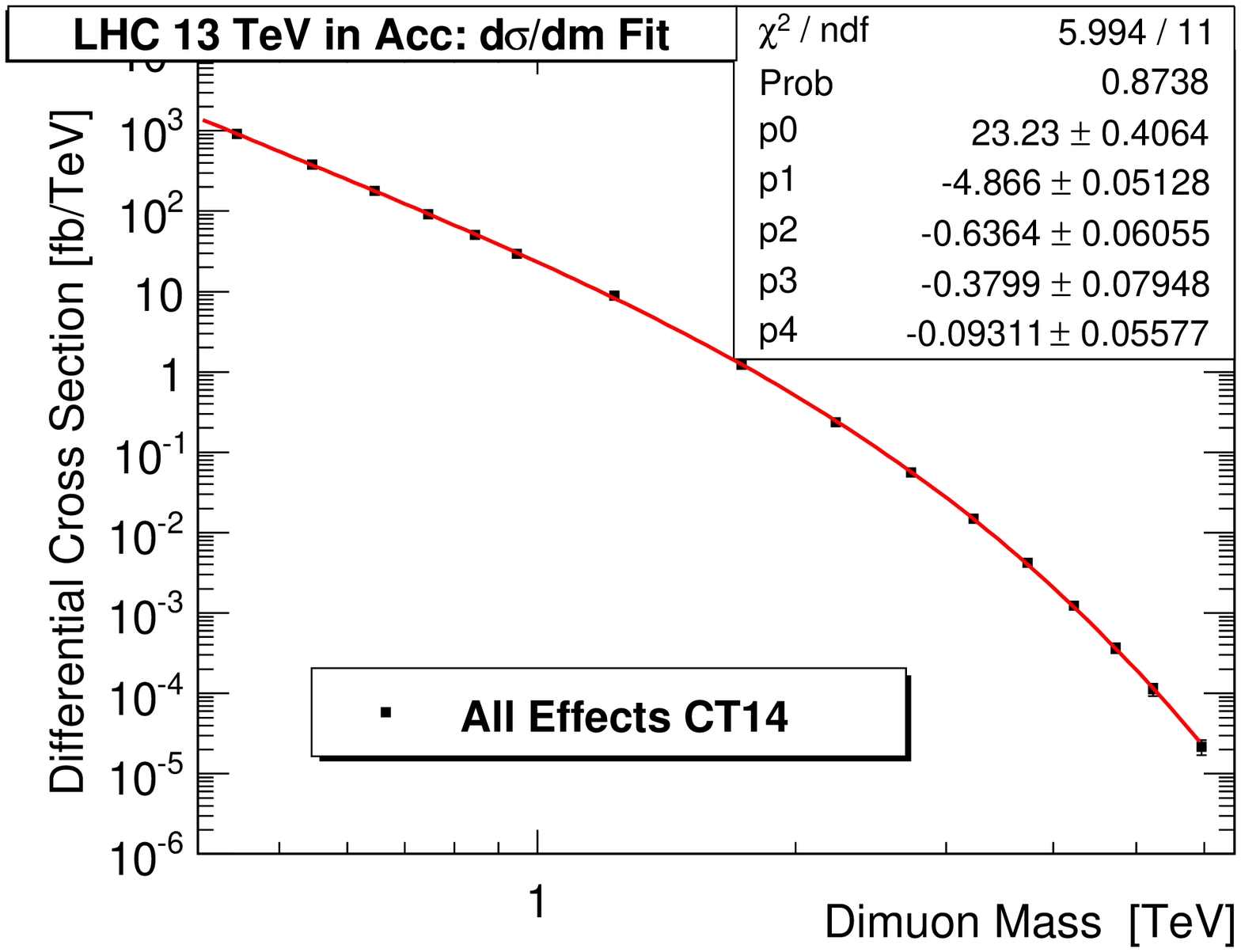}}}
\vspace*{-3pt}
\caption{Top: Fit to the differential cross section including all
         effects (NNLO and PI) for the the dielectron channel.
         Bottom: Fit to the differential cross section including all
         effects (NNLO and PI) for the dimuon channel.
         The CT14 PDF sets are used.}
\label{fig:fig8}
\end{figure}

Ad--hoc fits to the differential cross sections as function of mass
are shown in Figure~\ref{fig:fig8}. The following function is used:
\begin{equation}
\frac{d \sigma}{d m} = p_0\cdot m^{(p_1+p_2\cdot \ln m +p_3\cdot (\ln m)^2 +p_4\cdot (\ln m)^3)},
\end{equation}
where $p_0$ to $p_4$ are free parameters, m is the invariant mass of
the dilepton system in TeV, and $\sigma$ is the cross section in
fb. The fits, inspired by the ``not-too-far from linear'' behavior of
the curves on a double logarithmic plot, perform well if five
parameters are used. The $\chi^2$/d.o.f. is good, and the fit
parameters are shown on the plots for the two channels. The shapes of
the two distributions are identical within the statistical errors,
while the yield (compare the values of the $p_0$ parameter) is higher
in the electron channel, as discussed in the next paragraph.

\begin{figure}[ht]
\centerline{\resizebox{0.88\textwidth}{10.5cm}{\includegraphics{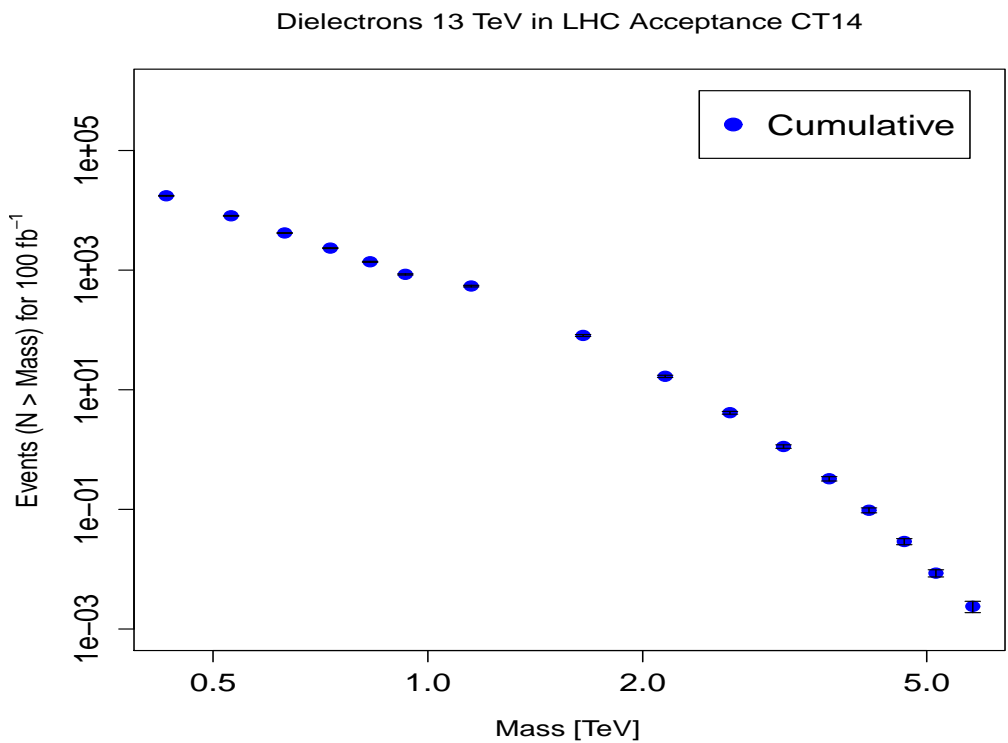}}}
\centerline{\resizebox{0.88\textwidth}{10.5cm}{\includegraphics{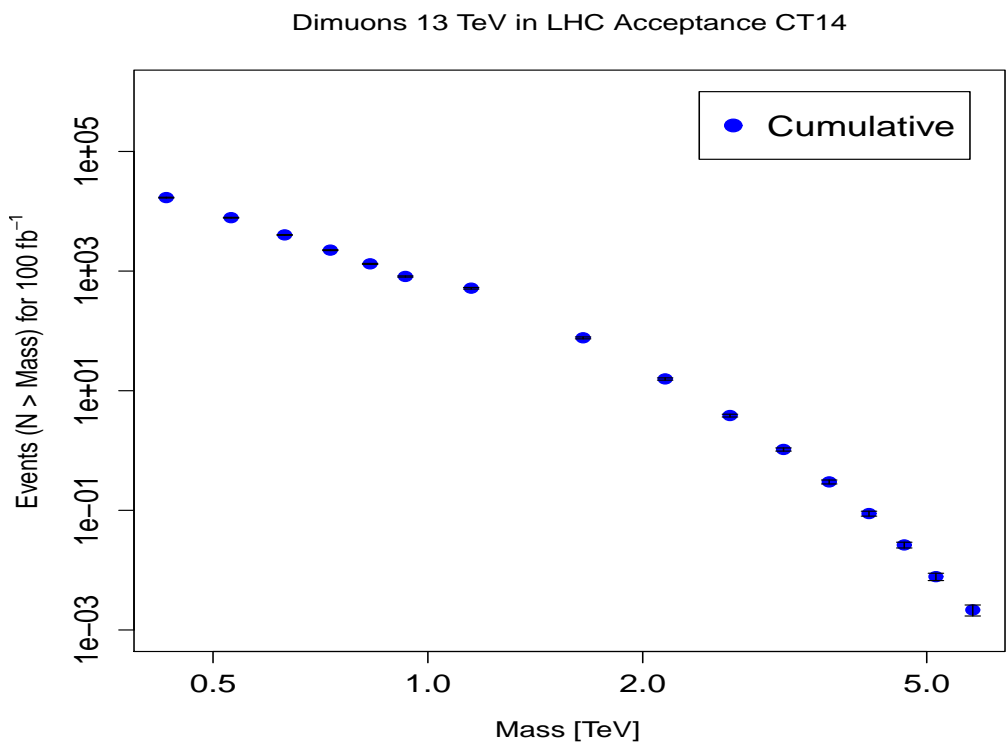}}}
\vspace*{-3pt}
\caption{Top: Cumulative number of events expected in one experiment above
         a given mass for integrated luminosity of 100~fb$^{-1}$ including
         all effects (NNLO and PI) for the dielectron channel.
         Bottom: Cumulative number of events expected in one experiment above
         a given mass for integrated luminosity of 100~fb$^{-1}$ including
         all effects (NNLO and PI) for the dimuon channel.}
\label{fig:fig9}
\end{figure}

The cumulative numbers of events, expected in one experiment above a
given mass for integrated luminosity of 100~fb$^{-1}$ at 13~$\TeV$,
are given in Figure~\ref{fig:fig9} for the two channels. The apparent
``kink'' in the distributions at 1 $\TeV$ is just an artifact of the
change of binning from 0.1 to 0.5 TeV bins at this point. The expected
numbers of events are summarized in Table~\ref{tab:tab1}. The yield
is slightly higher in the dielectron channel due to the recovery of
FSR radiation by the electromagnetic calorimeters. As a result fewer
events migrate to lower masses. In practice, this channel has an
additional advantage due to the favorable energy dependence of the
mass resolution. In the dimuon channels the mass resolution relies on
tracking and deteriorates at high mass. All things being equal the
dielectron channel might be first in a discovery. Additional search
options are provided by measuring the forward-backward asymmetry. Here
the muon channel may have an advantage, as the charge determination
for electrons relies on the central trackers with much shorter lever
arm, so it becomes increasingly difficult at high energies. Around one
event with mass exceeding 3 $\TeV$ per channel is expected for
luminosity of 100 fb$^{-1}$. Given the luminosities being collected by
ATLAS and CMS in 2016, with a bit of luck we may expect to see first
event(s) at these high masses this year.

\begin{table}[thb]
\renewcommand{\arraystretch}{1.10}
\caption{Cumulative expected numbers of events in one experiment above a
given mass for integrated luminosity of 100~fb$^{-1}$ at 13~$\TeV$,
using the CT14 PDF sets. The ``From Fit'' columns are obtained by
integrating the fits to the differential cross sections, as explained in
the text. All effects (NNLO and PI) are included.}
\vspace{0.3cm}
\begin{center}
{\begin{tabular}{|c||rll|r||rll|r|} \hline
  Mass & \multicolumn{4}{c||}{Dielectrons CT14}& \multicolumn{4}{c|}{Dimuons CT14} \\ \cline{2-9}
($\TeV$)&  Events  & Error$^+$& Error$^-$& From Fit &  Events  & Error$^+$& Error$^-$& From Fit\\
\hline
  0.4  &    17500  &  215     &  282     &   17600  &   16950  &  210     &  274     &   17050 \\
  0.5  &     8100  &  99      &  131     &    8090  &    7820  &  97      &  127     &    7800 \\
  0.6  &     4180  &  53      &  69      &    4200  &    4030  &  51      &  67      &    4040 \\
  0.7  &     2340  &  32      &  42      &    2370  &    2240  &  31      &  40      &    2280 \\
  0.8  &     1380  &  22      &  28      &    1420  &    1320  &  21      &  27      &    1360 \\
  0.9  &      850  &  17      &  22      &     890  &     810  &  16      &  21      &     850 \\
  1.0  &      540  &  15      &  19      &     570  &     520  &  14      &  18      &     550 \\
  1.5  &     81.0  &  2.7     &  3.2     &    87.5  &    76.7  &  2.6     &  3.1     &    82.9 \\
  2.0  &     16.8  &  0.68    &  0.78    &    17.9  &    15.8  &  0.64    &  0.74    &    16.8 \\
  2.5  &      4.1  &  0.21    &  0.23    &     4.3  &     3.8  &  0.20    &  0.22    &     4.0 \\
  3.0  &      1.1  &  0.073   &  0.077   &     1.1  &     1.0  &  0.062   &  0.064   &     1.0 \\
  3.5  &     0.33  &  0.027   &  0.027   &    0.32  &    0.30  &  0.023   &  0.023   &    0.29 \\
  4.0  &    0.097  &  0.0091  &  0.0091  &   0.094  &   0.088  &  0.0084  &  0.0083  &   0.086 \\
  4.5  &    0.029  &  0.0034  &  0.0034  &   0.029  &   0.026  &  0.0029  &  0.0029  &   0.026 \\
  5.0  &   0.0086  &  0.0012  &  0.0012  &  0.0085  &  0.0078  &  0.0011  &  0.0011  &  0.0078 \\
  5.5  &   0.0024  &  0.0005  &  0.0005  &  0.0021  &  0.0022  &  0.0005  &  0.0005  &  0.0019 \\
\hline
\end{tabular}}
\end{center}
\label{tab:tab1}
\end{table}

The event yields are reproduced well by integrating the fits to the
differential cross sections, shown in Figure~\ref{fig:fig8}. The
``From Fit'' columns in Table~\ref{tab:tab1}, obtained this way, agree
with the yields produced by calculating directly from the cross
sections (the ``Events'' columns). All that is needed for predictions
in different binnings is a new integration of the fit functions.

The total number of expected events above 0.4 $\TeV$ for the two
channels in one experiment for 100 fb$^{-1}$ is 34450. It is
interesting to compare this number to the early
study~\cite{Haywood:1999qg} from 1999. The center-of-mass energy is 14
$\TeV$, and the selection is not exactly the same: $|\eta| < 2.5$ and
$p_T > 20\ \GeV$ for both leptons. The simulation is done at leading
order with {\tt PYTHIA 5.7} and the then available PDFs. Given all
these caveats, the 1999 prediction: 33000 expected events, is
surprisingly close to the new result~\footnote{For the sake of full
disclosure, the author of the present paper and the person who
produced these numbers back then are identical.}.

\begin{figure}[ht]
\centerline{\resizebox{0.88\textwidth}{10.5cm}{\includegraphics{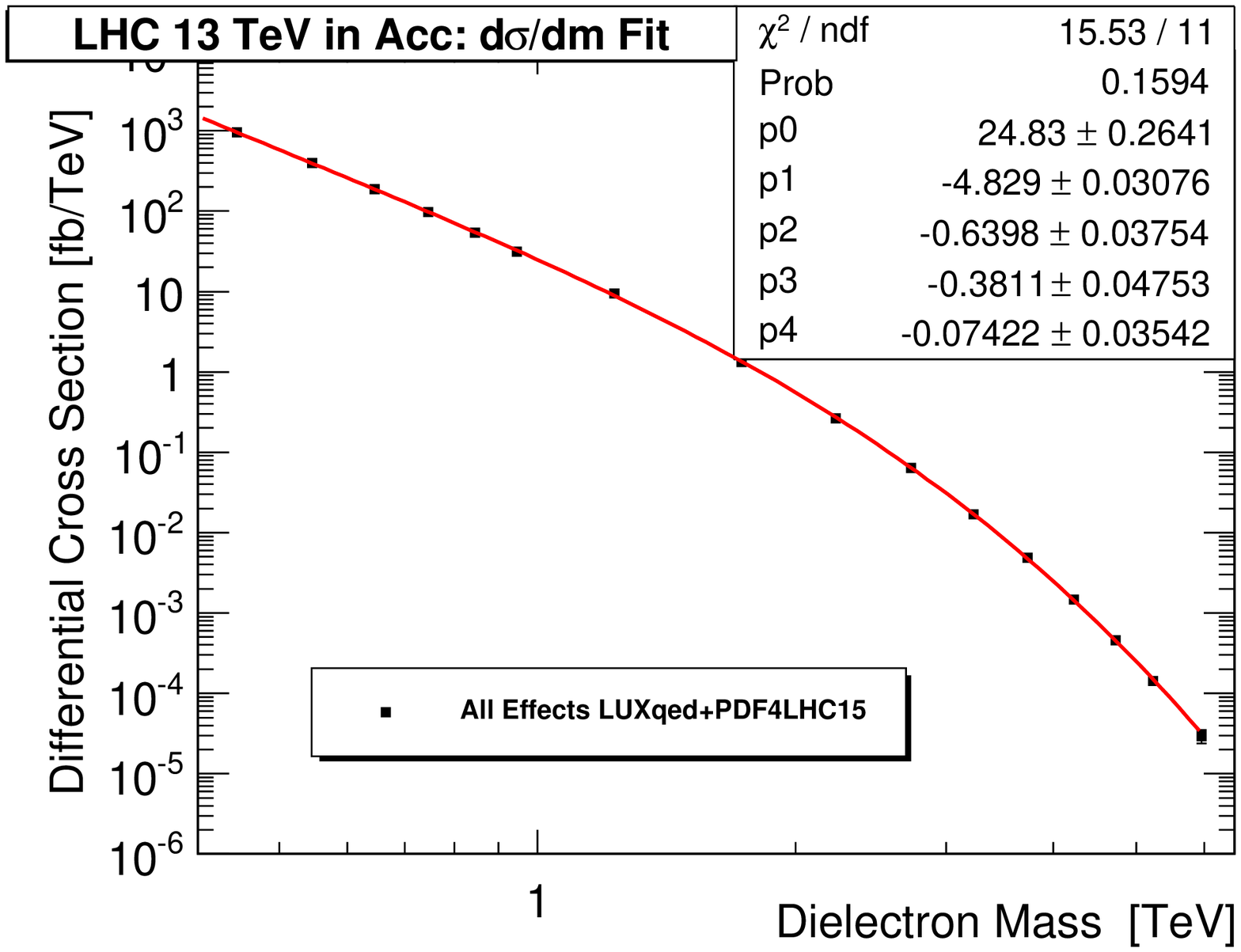}}}
\centerline{\resizebox{0.88\textwidth}{10.5cm}{\includegraphics{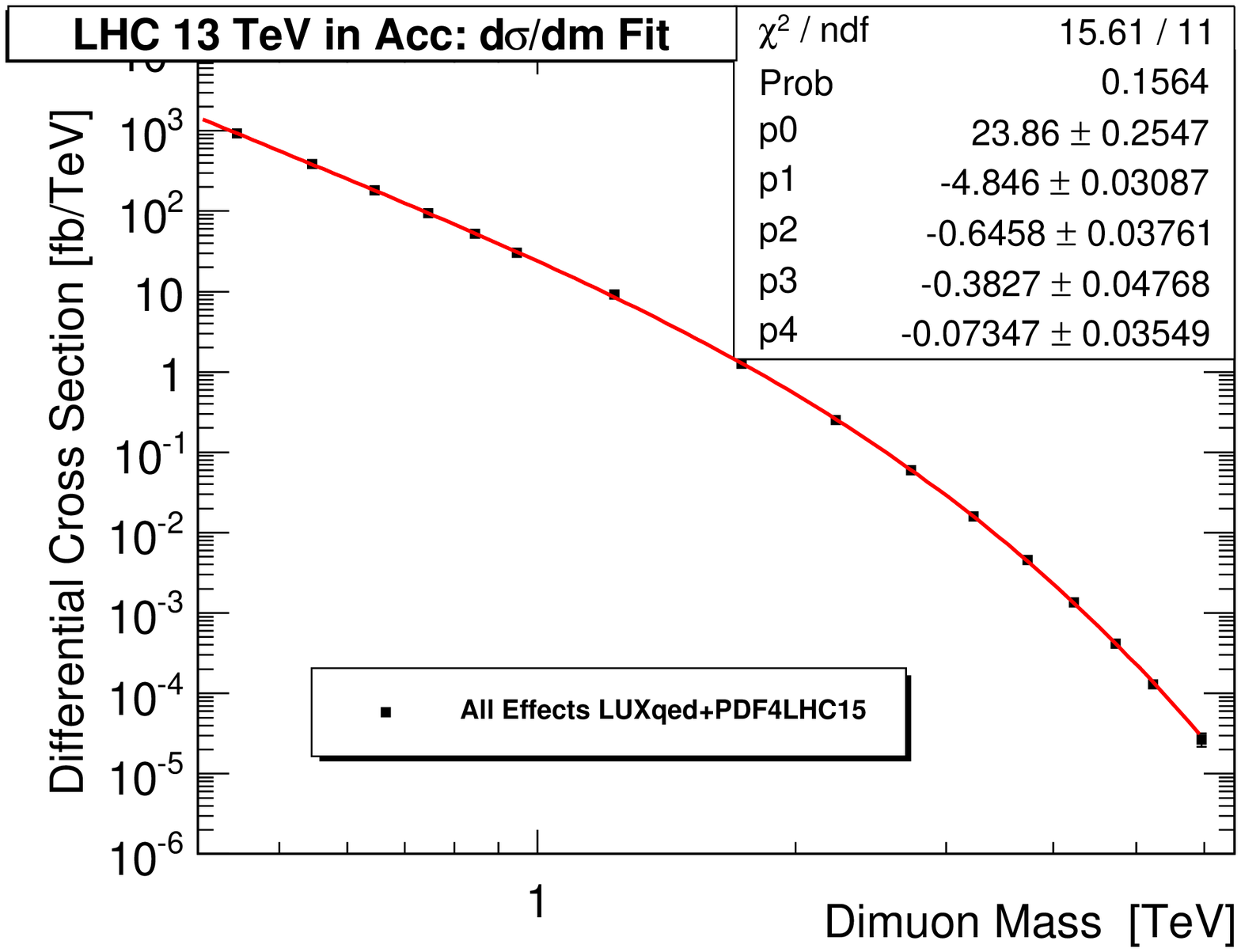}}}
\vspace*{-3pt}
\caption{Top: fit to the differential cross section including all
         effects (NNLO and PI) for the dielectron channel.
         Bottom: fit to the differential cross section including all
         effects (NNLO and PI) for the dimuon channel.
         The LUXqed\_plus\_PDF4LHC15 PDF set is used.}
\label{fig:fig8a}
\end{figure}

The cumulative numbers of events for the two channels, expected in one
experiment above a given mass for integrated luminosity of
100~fb$^{-1}$ at 13~$\TeV$, are given in Table~\ref{tab:tab2} using
the LUXqed\_plus\_PDF4LHC15\_nnlo\_100 set. This PDF set includes both
the NNLO and PI effects. As can be seen, the yields agree quite well
within the uncertainties with the yields in Table~\ref{tab:tab1} based
on the CT14 PDF sets. The LUXqed\_plus\_PDF4LHC15 set has smaller
overall uncertainties and tends to predict slightly higher rates, with
both the PI effects and the underlying NNLO PDF set contributing to
the enhancement. The event yields are reproduced well by integrating
the fit to the differential cross section, shown in
Figure~\ref{fig:fig8a}. The smaller uncertainties in this case result
in a somewhat larger $\chi^2$/d.o.f. Within errors the five parameters
from the fit agree with the results using CT14,
cf. Figure~\ref{fig:fig8}.

\begin{table}[thb]
\renewcommand{\arraystretch}{1.10}
\caption{Cumulative expected numbers of events in one experiment above a
given mass for integrated luminosity of 100~fb$^{-1}$ at 13~$\TeV$,
using the LUXqed\_plus\_PDF4LHC15 PDF set. The ``From Fit'' columns are
obtained by integrating the fits to the differential cross sections, as
explained in the text. All effects (NNLO and PI) are included.}
\vspace{0.3cm}
\begin{center}
{\begin{tabular}{|c||rll|r||rll|r|} \hline
  Mass & \multicolumn{4}{c||}{Dielectrons LUXqed\_plus\_PDF4LHC15}& \multicolumn{4}{c|}{Dimuons LUXqed\_plus\_PDF4LHC15} \\ \cline{2-9}
($\TeV$)&  Events  & Error$^+$& Error$^-$& From Fit &  Events  & Error$^+$& Error$^-$& From Fit\\
\hline
  0.4  &    17760  &  135     &  152     &   17870  &   17210  &  131     &  148     &   17330  \\
  0.5  &     8230  &  64      &  70      &    8220  &    7960  &  63      &  68      &    7940  \\
  0.6  &     4270  &  35      &  37      &    4280  &    4110  &  34      &  36      &    4130  \\
  0.7  &     2390  &  21      &  22      &    2420  &    2300  &  21      &  22      &    2330  \\
  0.8  &     1410  &  15      &  15      &    1450  &    1360  &  14      &  15      &    1400  \\
  0.9  &      870  &  12      &  12      &     910  &     840  &  11      &  11      &     870  \\
  1.0  &      560  &  10      &  10      &     590  &     530  &  9.8     &  9.7     &     560  \\
  1.5  &     84.3  &  1.9     &  1.7     &    90.7  &    80.0  &  1.8     &  1.6     &    86.1  \\
  2.0  &     17.6  &  0.46    &  0.40    &    18.7  &    16.6  &  0.43    &  0.37    &    17.6  \\
  2.5  &      4.4  &  0.13    &  0.11    &     4.5  &     4.1  &  0.12    &  0.11    &     4.2  \\
  3.0  &      1.2  &  0.039   &  0.037   &     1.2  &     1.1  &  0.036   &  0.034   &     1.1  \\
  3.5  &     0.35  &  0.013   &  0.014   &    0.35  &    0.33  &  0.012   &  0.013   &    0.32  \\
  4.0  &     0.11  &  0.0049  &  0.0054  &    0.11  &   0.098  &  0.0046  &  0.0049  &   0.097  \\
  4.5  &    0.033  &  0.0021  &  0.0023  &   0.033  &   0.030  &  0.0019  &  0.0021  &   0.030  \\
  5.0  &    0.010  &  0.0009  &  0.0011  &   0.010  &  0.0092  &  0.0009  &  0.0010  &  0.0092  \\
  5.5  &   0.0029  &  0.0005  &  0.0006  &  0.0026  &  0.0027  &  0.0005  &  0.0005  &  0.0023  \\
\hline
\end{tabular}}
\end{center}
\label{tab:tab2}
\end{table}

The most important test for the expected yields is the comparison to
the data from the LHC experiments. The ATLAS collaboration
observes~\cite{Aaboud:2016cth} 26 events above 0.9 $\TeV$ for
integrated luminosity 3.2 fb$^{-1}$. Corrections have to be applied to
this number to account for background contamination and detection
efficiency, which work in opposite directions. The ATLAS acceptance
extends to $|\eta| < 2.5$, but excludes the transition region $1.37 <
|\eta| < 1.52$ between the central and forward calorimeters. The
prediction from the numbers in Table~\ref{tab:tab1}, rescaled by
luminosity and pseudorapidity acceptance, is 26.6 events, in excellent
agreement with the data. The comparison in the muon channel is not so
straightforward due to the lower detection efficiency and higher
backgrounds. The updated preliminary ATLAS dielectron
numbers~\cite{ATLAS:2016cyf} for 13.3 fb$^{-1}$ are 99 observed
events. From the results presented here the expectation is 111
events, in good agreement with the data.

A 2.9 $\TeV$ event in the dielectron channel observed by
CMS~\cite{Khachatryan:2016zqb} is the highest mass event as of this
writing. From the numbers in Table~\ref{tab:tab1} and the luminosity
reported in the Run~2 papers cited here, the expectation is for
$\sim$~2 events above 2.5 $\TeV$ to be observed when combining the
ATLAS and CMS yields in the dielectron and dimuon channels.

\section{Outlook}

The production of high invariant mass opposite sign lepton pairs in
proton-proton collisions at the LHC is an important search region for
manifestations of new physics, and for tests of the Standard Model at
highest momentum transfers. In this paper the Drell-Yan and
photon-induced backgrounds for dilepton searches are examined in great
detail. Electroweak corrections, PDF uncertainties and choice,
$\alpha_s$ and scale dependencies, and QCD effects at
next-to-next-to-leading order are considered. The Drell-Yan background
is dominating at high masses, and the major source of uncertainty
comes from the limited knowledge of parton density functions in this
kinematic area. The photon-induced background plays a supporting
role. The backgrounds are low and well understood in the most
promising search region, and the LHC accelerator and the experiments
are delivering and recording record luminosities. The hunt is on.

\vskip .5cm
\noindent
{\large{\bf{Acknowledgments}}}
\noindent

The author thanks Valery Khoze for very productive discussions.


\clearpage

\clearpage

\clearpage

\clearpage

\clearpage

\clearpage

\clearpage

\clearpage

\clearpage

\clearpage

\clearpage

\clearpage

\end{document}